\documentclass[%
reprint,
superscriptaddress,
amsmath,amssymb,
pra,]{revtex4-2}
\usepackage[utf8]{inputenc}
\usepackage{float}
\makeatletter
\let\newfloat\newfloat@ltx
\makeatother

\usepackage{graphicx} 
\usepackage{graphicx,booktabs,array}
\usepackage[margin=1in]{geometry}
\usepackage[skip=2pt]{caption} 
\usepackage{subcaption}
\usepackage{dcolumn}
\usepackage{bm}
\usepackage{hyperref}
\usepackage{algorithm}
\usepackage{algpseudocode}
\usepackage{braket}
\usepackage{booktabs}
\usepackage{multirow}
\usepackage{tabularx}
\usepackage{siunitx}
\usepackage{orcidlink}
\DeclareSIUnit\angstrom{\AA}

\def\equationautorefname~#1\null{Eq.~(#1)\null}

\begin{document}

\title{Evaluating Ground State Energies of Chemical Systems with Low-Depth Quantum Circuits and High Accuracy}

\author{Shuo Sun\,\orcidlink{0009-0006-5775-9730}}
\email{shuo.sun@tum.de}
\affiliation{Technical University of Munich, School of Computation, Information and Technology, Boltzmannstra{\ss}e 3, 85748 Garching, Germany}
\author{Chandan Kumar\,\orcidlink{0000-0001-6510-4204}}
\affiliation{BMW Group Central Invention, 80788 Munich, Germany}
\author{Kevin Shen\,\orcidlink{0009-0005-2506-4056}}
\affiliation{BMW Group Central Invention, 80788 Munich, Germany}
\affiliation{applied Quantum algorithms (aQa), Leiden University, Leiden, The Netherlands}
\author{Elvira Shishenina}
\affiliation{BMW Group Central Invention, 80788 Munich, Germany}
\author{Christian B. Mendl\,\orcidlink{0000-0002-6386-0230}}
\email{christian.mendl@tum.de}
\affiliation{Technical University of Munich, School of Computation, Information and Technology, Boltzmannstra{\ss}e 3, 85748 Garching, Germany}
\affiliation{Technical University of Munich, Institute for Advanced Study, Lichtenbergstra{\ss}e 2a, 85748 Garching, Germany}

\date{\today}

\begin{abstract}
Solving electronic structure problems is considered one of the most promising applications of quantum computing. However, due to limitations imposed by the coherence time of qubits in the Noisy Intermediate Scale Quantum~(NISQ) era or the capabilities of early fault-tolerant quantum devices, it is vital to design algorithms with low-depth circuits.
In this work, we develop an enhanced Variational Quantum Eigensolver (VQE) ansatz based on the Qubit Coupled Cluster~(QCC) approach, which demands optimization over only $n$ parameters rather than the usual $n+2m$ parameters, where $n$ represents the number of Pauli string time evolution gates $e^{-itP}$, and $m$ is the number of qubits involved. 
We evaluate the ground state energies of $\mathrm{O_3}$, $\mathrm{Li_4}$, and $\mathrm{Cr_2}$, using CAS(2,2), (4,4) and (6,6) respectively in conjunction with our enhanced QCC ansatz, UCCSD (Unitary Coupled Cluster Single Double) ansatz, and canonical CCSD method as the active space solver, and compare with CASCI results.
Finally, we assess our enhanced QCC ansatz on two distinct quantum hardware, IBM Kolkata and Quantinuum H1-1.
\begin{description}
\item[Keyword]
electronic structure problem, variational quantum eigensolver, qubit coupled cluster
\end{description}
\end{abstract}

\maketitle
\section{Introduction}
\label{sec: introduction}

Solving electronic structure problems is the cornerstone of computational chemistry, enabling us to unravel the dynamic and kinetic properties of chemical systems. 
In recent years, quantum computing has emerged as a promising avenue for efficiently simulating quantum systems. The Variational Quantum Eigensolver (VQE), which is first developed by Peruzzo et al.~\cite{peruzzo_variational_2014}, stands out as a leading algorithm for near-term quantum devices due to the shallow circuit and noise resiliency~\cite{tilly_variational_2022, sharma_noise_2020, fontana_evaluating_2021, huggins_efficient_2021}. VQE exploits the variational nature of the ground state energy, i.e.,
$\bra {\Phi_\text{init}} U(\mathbf{\theta})^\dagger H U(\mathbf{\theta}) \ket{ \Phi_\text{init} } \geq E_0,$
where $\ket{\Phi_\text{init}}$ is the initial state, $U(\mathbf{\theta})$ is parameterized unitary gates, which is often referred as ``ansatz", and $H$ is the Hamiltonian. Under Born-Oppenheimer approximation, the electronic Hamiltonian of a molecular system is given by
\begin{align}
	\hat{H}_{\text{el}} = \sum_{pq} h_{pq} \hat{a}_p^\dagger \hat{a}_q + \frac{1}{2} \sum_{pqrs} h_{pqrs} \hat{a}_p^\dagger \hat{a}_q^\dagger \hat{a}_s \hat{a}_r
\end{align}
in the second quantization form, where $p$, $q$, $r$ and $s$ are spin orbitals.
Various ans\"atze have been explored within VQE, such as Unitary Coupled Cluster~(UCC)~\cite{bartlett_alternative_1989, romero_strategies_2018, filip_stochastic_2020, metcalf_resource-efficient_2020, anand_quantum_2022}, Qubit Coupled Cluster~(QCC)~\cite{ryabinkin_qubit_2018, ryabinkin_iterative_2020, yordanov_qubit-excitation-based_2021}, the Hardware Efficient Ansatz~(HEA)~\cite{kandala_hardware-efficient_2017, mitarai_generalization_2019}, and the Adaptive ansatz~\cite{grimsley_adaptive_2019, claudino_benchmarking_2020, zhang_mutual_2021, tang_qubit-adapt-vqe_2021, grimsley_adapt-vqe_2023}, each tailored to capture specific features of the electronic structure. 

To achieve the global minimum, namely the ground state energy, an optimization loop is essential. However, many aforementioned ans\"atze require a plethora of parameters, deepening the circuits and making the optimization process computationally intensive. Additionally, estimating the gradient often demands a significant number of circuit executions, further exacerbating the computational load~\cite{crooks_gradients_2019, ostaszewski_structure_2021, izmaylov_analytic_2021}. In this article, we address the challenges posed by the large number of parameters and deep circuits by an enhanced QCC approach. In contrast to the original QCC ansatz, our enhanced approach starts from a Hartree-Fock state, ensuring the correct particle number from the beginning.
In this way, we eliminate the need for a series of single-qubit rotation gates in our enhanced approach to adjust the total particle number, as was the case in the original QCC ansatz. Then a sequence of Pauli string time evolution gates $e^{-it{P}}$, where ${P}$ is a tensor product of single-qubit Pauli operators, are applied. 

As demonstration, we assess the ground state energies and the potential energy surfaces of three molecules, $\mathrm{O_3}$, $\mathrm{Li_4}$, and $\mathrm{Cr_2}$, with the enhanced QCC ansatz, UCCSD ansatz, and CCSD as the active space solvers, and compare with CASCI results. We conduct a comparative analysis of the performance and the number of parameters required by the enhanced QCC and UCCSD ansatz. Furthermore, we execute the parametrized QCC circuits on two distinct quantum computers and achieve near chemical accuracy on one of the machines, which highlights the practical utility of our approach~\cite{kandala_hardware-efficient_2017, tilly_reduced_2021, bentellis_benchmarking_2023}.

For ease of notation, we follow the notation from~\cite{helgaker_molecular_2014} in this article, where occupied orbitals are denoted as $ab$, virtual (unoccupied) orbitals as $mn$, inactive orbitals as $ijkl$, active orbitals as $uvxy$, and general orbitals as $pqrs$. We decorate operators with $\space \hat{} \space$ to represent fermionic operators; the ones without decoration are qubit operators.

\section{Methods}
\label{sec: methods}

\subsection{Unitary Coupled Cluster}
\label{subsec: ucc}

The inspiration for UCC ansatz stems from the Coupled Cluster~(CC) theory in computational chemistry~\cite{bartlett_alternative_1989}. The latter approach involves applying cluster operators to a reference state, typically the Hartree-Fock (HF) state, resulting in a linear combination of Slater determinants with various excitations. First, let us define the excitation operators $\hat{T}$~\cite{anand_quantum_2022}:
\begin{subequations}
	\begin{align}
		\hat{T} &= \sum_{k=1}^n \hat{T}_k, \\
		\hat{T}_k &= \frac{1}{(k!)^2} \sum_{ab\dots}^{n_\text{occ}}\sum_{mn\dots}^{n_\text{vir}} t_{ab\dots}^{mn\dots} \hat{\tau}_{ab\dots}^{mn\dots}, \\
		\hat{\tau}_{ab\dots}^{mn\dots} &= \hat{a}_m^\dagger \hat{a}_n^\dagger \cdots \hat{a}_b \hat{a}_a.
	\end{align}
\end{subequations}
In the case where only single and double excitations are considered, the excitation operator simplifies to:
\begin{equation}
	\hat{T}_\mathrm{SD} = \sum_{a}^{n_\text{occ}}\sum_{m}^{n_\text{vir}} t_a^m \hat{a}^\dagger_m \hat{a}_a + \sum_{a>b}^{n_\text{occ}}\sum_{m>n}^{n_\text{vir}} t_{ab}^{mn} \hat{a}^\dagger_m \hat{a}^\dagger_n \hat{a}_b \hat{a}_a.
\end{equation}
For the canonical CCSD method, the cluster operator is given by $\mathrm{e}^{\hat{T}_\mathrm{SD}}$. However, this operator is not unitary in general and cannot be directly implemented on a quantum computer. Recognizing that $\hat{T} - \hat{T}^\dagger$ is a skew-hermitian operator, we instead apply $\mathrm{e}^{\hat{T} - \hat{T}^\dagger}$ as the unitary cluster operator, which is suitable for quantum computers. 

Since all the aforementioned operators are fermionic, fermion-to-qubit mappings such as Jordan-Wigner~\cite{jordan_uber_1928}, Bravi-Kitaev~\cite{bravyi_fermionic_2002}, Parity mapping~\cite{seeley_bravyi-kitaev_2012}, etc. are required to transform them into qubit operators. In the context of UCCSD, the resultant qubit operators often turn out too intricate for direct implementation on quantum computers. Consequently, an additional decomposition is essential, leading to more complex circuits in practical applications~\cite{grimsley_is_2020}.

\subsection{Qubit Coupled Cluster}
\label{subsec: qcc}

The QCC ansatz is both inspired by chemistry and designed for hardware efficiency. This approach uses a sequence of multi-qubit Pauli string time evolution gates defined as
\begin{equation}
	U(\tau) = \prod_{j=1}^{n} \mathrm{e}^{-i\tau_j {P}_j/2},
\end{equation}
where $\tau_j$ represents the coefficient and ${P}_j$ is the multi-qubit Pauli string operator. These gates act on an unentangled, qubit-mean-field~(QMF) state~\cite{ryabinkin_qubit_2018, ryabinkin_iterative_2020}, which is expressed as a tensor product of qubit states with each qubit state defined as:
\begin{align}
	\ket{\mathbf{\Omega}}  &= \bigotimes_{i=1}^{m} \ket{\Omega_i }, \\
	\ket{\Omega_i } &= \cos \left(\frac{\theta_i}{2}\right)\ket{0} + \mathrm{e}^{i\phi_i} \sin \left(\frac{\theta_i}{2}\right)\ket{1}.\label{eq: qmf}
\end{align}
The QCC energy, 
\begin{equation}
	E_{\text{QCC}} = \min_{\mathbf{\Omega},\mathbf{\tau}}\bra{ \mathbf{\Omega}} U(\mathbf{\tau})^\dagger {H} U(\mathbf{\tau})\ket{\mathbf{\Omega}}, \label{eq: optimise qcc}
\end{equation}
is then minimized over the parameters $\theta$s, $\phi$s and $\tau$s. 
Let us write the molecular Hamiltonian as sum of Pauli string operators 
\begin{equation}
	{H} = \sum_{k=1}^{M} C_k {P}_k.
\end{equation}
As we know that the electronic Hamiltonian is real when no external electric field is present~\cite{szabo1996modern}, the energy gradient with respect to $\tau_j$ can be simplified to~(see \cite{ryabinkin_iterative_2020} for details):
\begin{align}
	\frac{\partial E_{\mathrm{QCC}}[{P}_j]}{\partial \tau_j} &= \sum_k C_k \mathrm{Im} \bra{ \mathbf{\Omega}_{\min}} {P}_k {P}_j \ket{\mathbf{\Omega}_{\min}},
\end{align}
where $\ket{\mathbf{\Omega}_{\min}}$ is the QMF state with the lowest energy, i.e.,
$\ket{\mathbf{\Omega}_{\min}} = \mathrm{arg} \min \bra{\mathbf{\Omega}} {H} \ket{\mathbf{\Omega}}$.
With the assumption that the QMF state $\ket{\mathbf{\Omega}_{\min}}$ is an eigenstate of all $\{\hat{Z_i}\}_{i=1}^{n_q}$ operators, we can further reduce the computational cost by grouping the Pauli terms in the Hamiltonian that share the same flip index $F({P})$ together, where 
\begin{equation}
	F({P}) = \big\{i:{P}^i \in \{{X},{Y}\} \big\},
	\label{eq: flip index}
\end{equation}
as they exhibit gradients with the same absolute value.

After selecting the Pauli strings ${P_j}$s with the highest gradients, and optimizing the corresponding amplitudes $\tau_i$, we proceed to the next iteration. The Hamiltonian is updated by
\begin{align}
	{H}^{(r+1)} &= \Bigg(\prod_{j=1}^{n}\mathrm{e}^{-i\tau_j {P}_j/2}\Bigg)^\dagger {H}^{(r)} \Bigg(\prod_{j=1}^{n}\mathrm{e}^{-i\tau_j {P}_j/2} \Bigg).
	\label{eq: dress hamiltonian}
\end{align}
Here $n$ is the number of Pauli strings that are applied in each iteration, and $r$ is the iteration index.

One caveat of QCC is that the parameters for the initial state scale linearly with the system size, and optimization challenges, such as the Barren plateaus~\cite{mcclean_barren_2018, wang_noise-induced_2021}, can arise.

\subsection{Our contribution to QCC}
\label{subsec: our contribution to qcc}

In this work, we aim to develop an ansatz that uses low-depth circuits and fewer parameters. A key observation driving our approach is the non-conservation of the particle number inherent in arbitrary Pauli time evolution gates, i.e., $[N,e^{-i\tau {P}/2}] \neq 0$. However, in the context of a chemical system, the ground state inherently possesses a well-defined particle number. Moreover, in a chemical system with a stable configuration, a state with a different particle number incurs higher energy than a state with the correct particle number. To address this, the original ansatz introduces additional single-qubit rotation gates at its onset, as shown in~\autoref{eq: qmf}, to guide the state to the ground state particle number sector~\cite{particle-number}.

Our innovation comes from using the state with the correct particle number, like HF state, as the initial state, followed by applying a series of Pauli string time evolution gates $e^{-i\tau {P}/2}$. In this way, we can eliminate the need for single-qubit rotation gates. Additionally, it's noteworthy that in the molecular orbital basis, the HF state satisfies the assumption made in the original ansatz: it is indeed an eigenstate of all $\{Z_i\}_{i=1}^{n_q}$ operators. As a result, we can inherit the remainder of the original QCC ansatz.

For a problem with $m$ system qubits, preparing a QMF state, as defined in~\autoref{eq: qmf}, requires $2m$ single-qubit rotation gates and parameters, along with $n$ Pauli string time evolution gates for evolving the system towards gradient descent. In contrast, our enhanced approach allows the reduction of single-qubit rotation gates, resulting in a significant decrease in both the number of parameters and computational demand. A detailed algorithmic procedure is presented in~\autoref{alg: enhanced iqcc}.

\begin{algorithm}[ht]
	\caption{Enhanced QCC}\label{alg: enhanced iqcc}
	\begin{algorithmic}[1]
		\Require Fermionic molecular Hamiltonian: $\hat{H}_\text{mol}^f$, Hartree-Fock state: $\ket{\Psi_\text{HF}}$, maximal iteration: \texttt{maxiter}, convergence tolerance: $\epsilon_{\text{tol}}$, number of Pauli strings in each iteration: $n$
		\Ensure $E = \min_{\theta} \bra{\Psi_\text{HF}}U(\theta)^\dagger H U(\theta) \ket{\Psi_\text{HF}}$
		\vspace{0.3cm}
		\State $H_\text{mol}^{q} \gets$ \texttt{FermionToQubitMapping}$(\hat{H}_{\text{mol}}^{f})$
		\State $\ket{\Psi_\text{init}}\gets$ \texttt{FermionToQubitMapping}$(\ket{\Psi_\text{HF}})$
		\While{$n_\text{iter} <$ \texttt{maxiter}}
		\While{$\epsilon>\epsilon_\text{tol}$}
		\State $\text{paulis} \gets$ \texttt{SelectPauliString}($H_\text{mol}^{q}$, $n_g$)\Comment{\autoref{eq: flip index}}  
		\State $E, \theta \gets$ \texttt{Optimise}($\text{paulis}, H_\text{mol}^q$)\Comment{\autoref{eq: optimise qcc}}
		\State $H_\text{mol}^q \gets$ \texttt{DressHamiltonian}($\text{paulis}, H_\text{mol}^q, \theta $)\Comment{\autoref{eq: dress hamiltonian}}
		\State $\epsilon = E_\text{old}-E_\text{new}, n_\text{iter}+=1, E_\text{old}=E$
		\EndWhile
		\EndWhile
	\end{algorithmic}
\end{algorithm}

\subsection{Complete Active Space}
\label{subsec: active space}

The Complete Active Space (CAS) approach is a powerful tool in computational chemistry, reducing the problem size while maintaining promising accuracy for the total energy calculation. 
Constrained by the available computing resources, we use CAS to reduce the problem size in the examples presented in~\autoref{sec: numerical results}.

The active space Hamiltonian, denoted as $\hat{H}_\text{active}$, is expressed as~\cite{rossmannek_quantum_2021}:
\begin{align}
	\hat{H}_{\text{active}} = \sum_{uv} F_{uv}^{I} \hat{a}_u^\dagger \hat{a}_v + \sum_{uvxy} g_{uvxy} \hat{a}_u^\dagger \hat{a}_v^\dagger \hat{a}_y \hat{a}_x,
\end{align}
where $F_{uv}^I$ is defined as:
\begin{align}
	F_{uv}^I = h_{uv} + \sum_{i} (2 g_{iiuv} - g_{ivui})
\end{align}
and the inactive space energy is
\begin{align}
	E^\text{inactive} = \frac{1}{2} \sum_{ij} (h_{ij} + F_{ij}^I)D_{ij}^I.
\end{align}

The one-body term in the Hamiltonian is updated to describe not only the kinetic energy of the active space electrons but also the energy contributed by the inactive electrons. The two-body term consists of slices of the original two-body interaction. The total energy ($E^\text{total}$) is then expressed as the sum of the active space energy ($E^\text{active}$), the inactive space energy ($E^\text{inactive}$), and the energy contributed by the nuclei ($E^\text{nuclei}$):
\begin{align}
	E^\text{total}=E^\text{active}+E^\text{inactive}+E^\text{nuclei}.
\end{align}

In this paper, we will use CAS$(e,o)$ to represent the active space size, where $e$ is the number of electrons and $o$ is the number of spatial orbitals.

\section{Numerical results}
\label{sec: numerical results}

\begin{figure}[ht]
	\centering
	\begin{subfigure}[t]{0.3\linewidth}
		\caption{}
		\includegraphics[width=\linewidth]{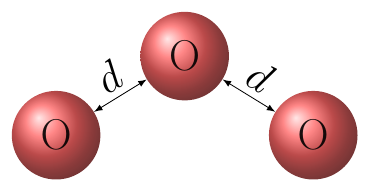}
		\label{fig: o3 molecule} 
	\end{subfigure}
	\begin{subfigure}[t]{0.3\linewidth}
		\caption{}
		\includegraphics[width=1.1\linewidth]{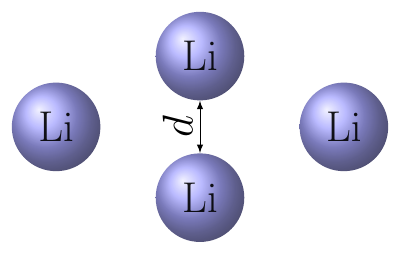}
		\label{fig: li4 molecule} 
	\end{subfigure}
	\begin{subfigure}[t]{0.3\linewidth}
		\caption{}
		\includegraphics[width=0.7\linewidth]{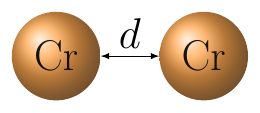}
		\label{fig: cr2 molecule} 
	\end{subfigure}
	\caption{Geometries of (a): $\mathrm{O_3}$ molecule, (b): $\mathrm{Li_4}$ molecule, and (c): $\mathrm{Cr_2}$ molecule. $d$ is the bond length we refer to throughout the paper.}
	\label{fig: molecule coordinates}
\end{figure}

Within this section, we numerically demonstrate our algorithm by computing the ground state energies and the potential energy surfaces of $\mathrm{O_3}$, $\mathrm{Li_4}$, and $\mathrm{Cr_2}$ molecules with active space size (2,2), (4,4), and (6,6). The geometry of the molecules are shown in~\autoref{fig: molecule coordinates}, and we utilize the cc-pvdz basis set~\cite{dunning_gaussian_1989} and Parity mapping~\cite{seeley_bravyi-kitaev_2012} for all calculations. 
We employed PySCF~\cite{sun_pyscf_2018} to obtain the active space Hamiltonians, CCSD, and CASCI results. Quantum circuit simulations were carried out using the noiseless statevector simulator in Qiskit~\cite{Qiskit}. 
The experimental results obtained from runs on quantum hardware are detailed in~\autoref{sec: experimental results}.

\subsection{Convergence}
\label{subsec: convergence}

\begin{figure}[ht]
	\centering
	\includegraphics[width=\linewidth]{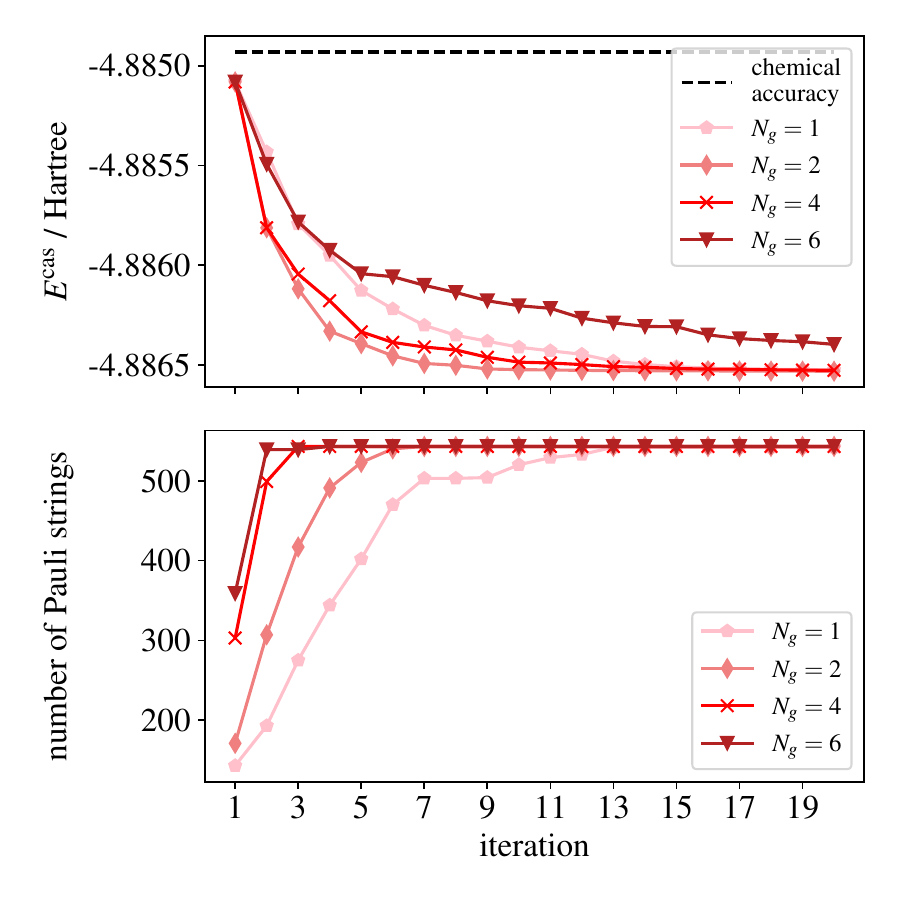}
	\caption{Energy convergence~(upper panel) and number of terms in the Hamiltonian~(lower panel) using enhanced QCC ansatz with different numbers of generators for $\mathrm{O_3}$ at $d=1.28$~\AA.}
	\label{fig: convergence total}
\end{figure}

We first focused on examining the convergence behavior of our algorithm, particularly in the context of ground state energy estimation for $\mathrm{O_3}$ and $\mathrm{Li_4}$ at two configurations. The representative result, illustrated in~\autoref{fig: convergence total}, revealed interesting patterns as we varied the number of generators. Notably, we found that altering the number of generators at each iteration did not have a significant impact on the convergence energy nor the number of terms in the final Hamiltonian. This observation suggested the robustness of our algorithm regardless of the specific generator count.

Moreover, as depicted in~\autoref{fig: convergence total}, an increase in the number of generators at each iteration exhibited a tendency to accelerate the convergence to some degree. Nevertheless, a critical threshold was observed beyond which the convergence rate diminished. This phenomenon is likely attributed to the exponential expansion of the search space with the growing number of generators, presenting a challenge for the optimizer to efficiently locate the optimal solution within a limited number of iterations.

Additionally, our investigation revealed that while the number of terms in the Hamiltonian saturated more rapidly with higher number of generators, this saturation did not necessarily correlate with the convergence of energy as shown in~\autoref{fig: convergence total}. For the sake of optimization simplicity, we opted to use only one generator per iteration in the subsequent examples.

\begin{figure*}
	\centering
	\includegraphics[width=0.9\linewidth]{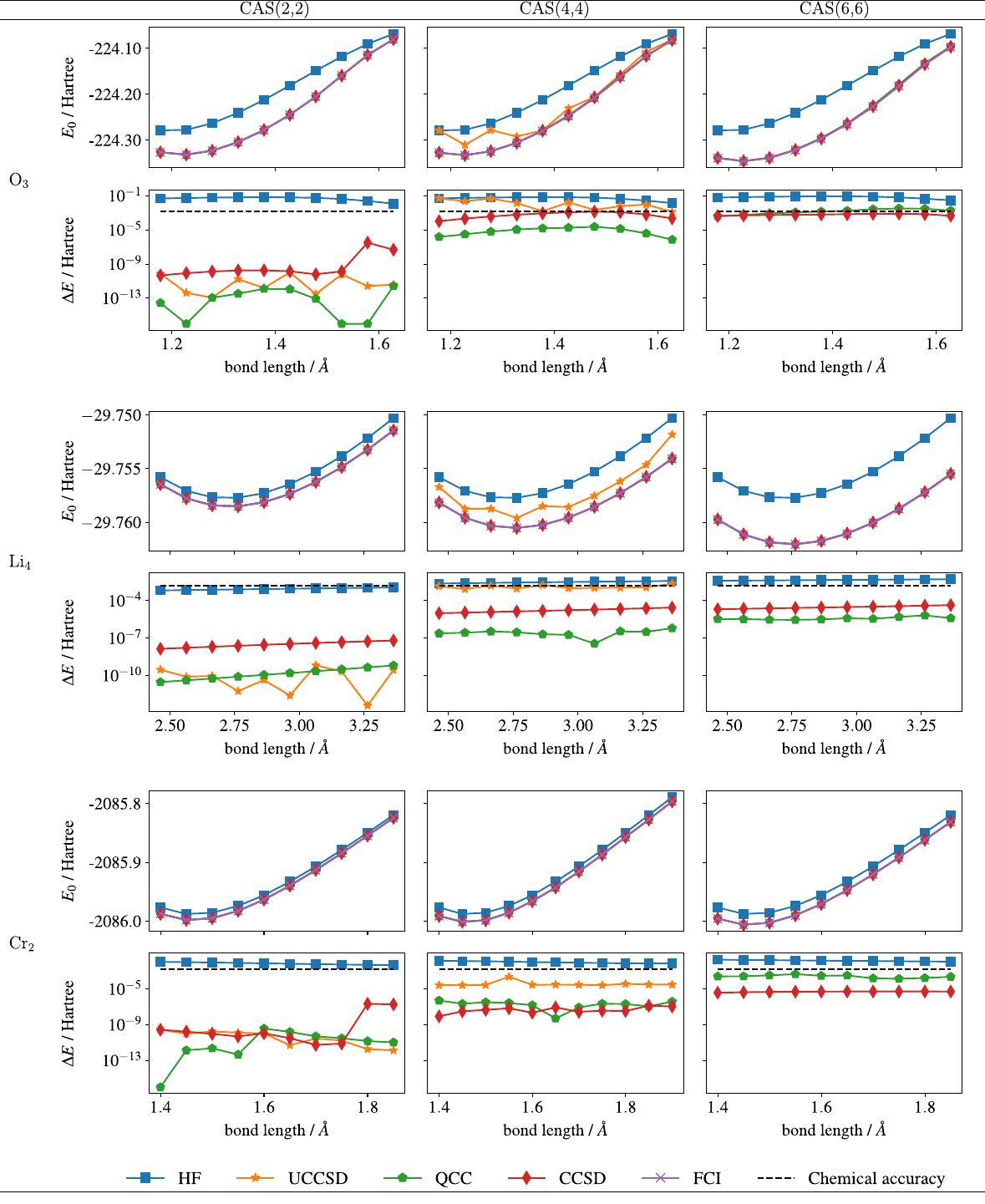}
	\caption{Potential energy surfaces of $\mathrm{O_3}$, $\mathrm{Li_4}$, and $\mathrm{Cr_2}$, using CAS(2,2), (4,4) and (6,6) in conjunction with the enhanced QCC ansatz, UCCSD (Unitary Coupled Cluster Single-Double) ansatz, CCSD, and FCI as the active space solver. The upper panel of each figure describes the ground state energy at different configurations. The lower panel of each figure shows the energy difference between the chosen active space solver and CASCI (CAS with FCI as active space solver). For active space (6,6), due to the high parameter counts of UCCSD method, we did not perform the PES calculation.}
	\label{fig:pes_total_table}
\end{figure*}

\subsection{Potential energy surfaces}
\label{subsec: pes}

We investigated potential energy surfaces for various molecules using CAS(2,2), CAS(4,4), and CAS(6,6) with four distinct active space solvers: UCCSD, QCC, CCSD (coupled cluster single and double), and FCI (full configuration interaction). The results are presented in~\autoref{fig:pes_total_table}.

First, we assessed the potential energy surface with active space of (2,2), i.e., two electrons in two spatial orbitals. In this case, all the active space solvers achieved chemical accuracy across all bond lengths for $\mathrm{O_3}$, $\mathrm{Li_4}$, and $\mathrm{Cr_2}$. 

However, when extending the active space to (4,4), the UCCSD solver faced challenges. It failed to produce energies within chemical accuracy and even returned energies at the Hartree-Fock level for $\mathrm{O_3}$ at specific bond lengths (e.g., $d=1.18$~\AA~and $1.28$~\AA). In contrast, the enhanced QCC solver consistently provided favorable results, occasionally surpassing the performance of the CCSD solver. This could arise from the inclusion of higher-order excitations in QCC, compared to the sole consideration of single and double excitations in CCSD.

\begin{figure*}[ht]
	\centering
	\begin{subfigure}[t]{0.495\linewidth}
		\caption{ }     
		\label{fig: (6,6) o3 convergence all}  
		\includegraphics[width=0.999\linewidth]{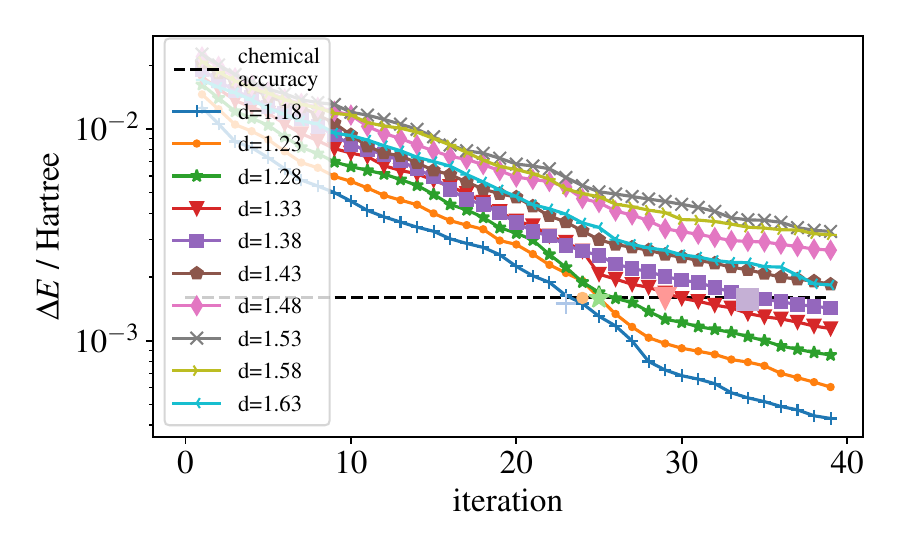}
	\end{subfigure}
	\begin{subfigure}[t]{0.495\linewidth}
		\caption{}
		\label{fig: (6,6) o3 fit 0.15}  
		\includegraphics[width=0.999\linewidth]{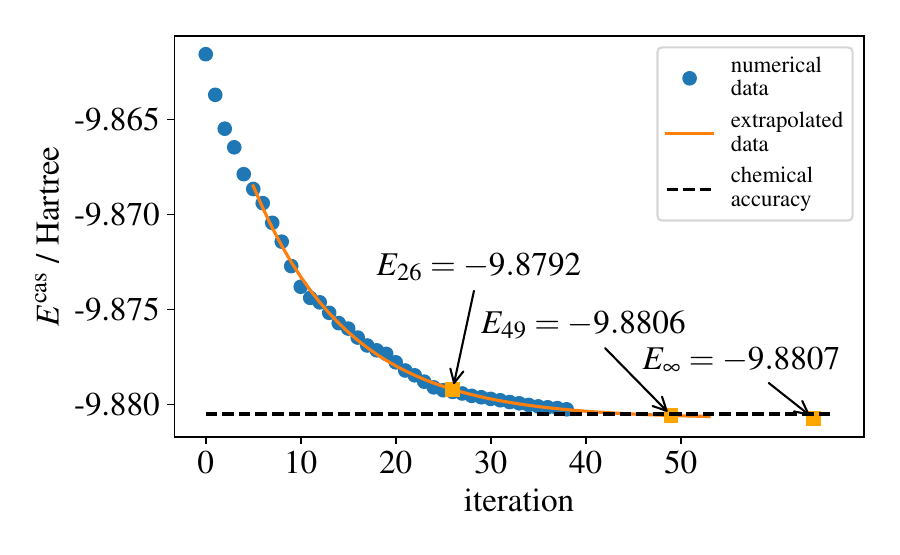}	    
	\end{subfigure}
	\vspace{-2mm}
	\begin{subfigure}[t]{0.495\linewidth}
		\caption{ }     
		\label{fig: (6,6) o3 fit 0.20}  
		\includegraphics[width=0.999\linewidth]{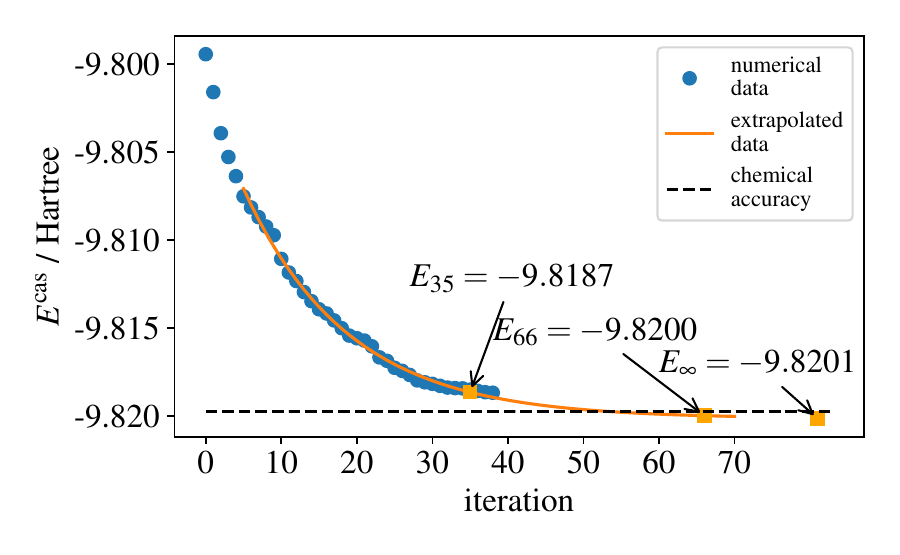}
	\end{subfigure}
	\begin{subfigure}[t]{0.495\linewidth}
		\caption{}
		\label{fig: (6,6) o3 fit 0.25}  
		\includegraphics[width=0.999\linewidth]{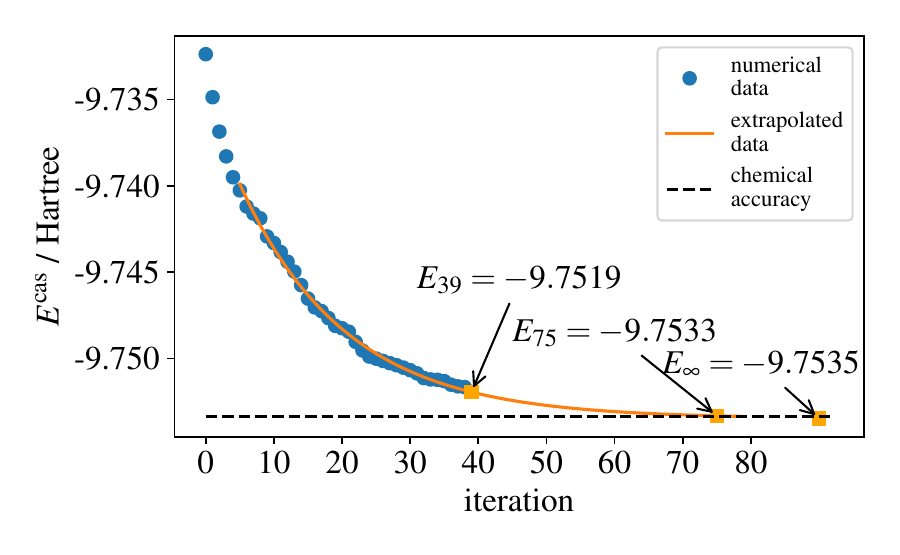}	     
	\end{subfigure}
	\vspace{-2mm}
	\begin{subfigure}[t]{0.495\linewidth}
		\caption{ }     
		\label{fig: (6,6) o3 fit 0.30}  
		\includegraphics[width=0.999\linewidth]{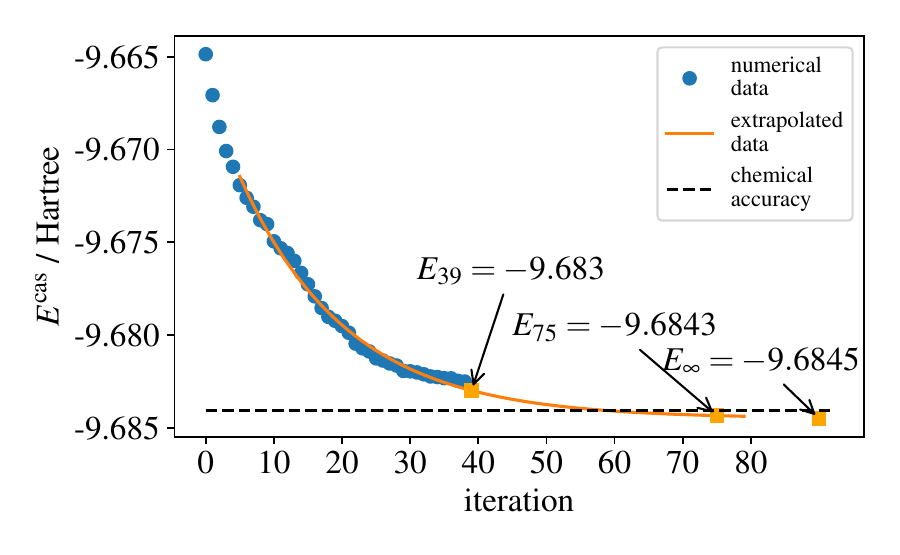}
	\end{subfigure}
	\begin{subfigure}[t]{0.495\linewidth}
		\caption{ }     
		\label{fig: (6,6) o3 fit 0.35}  
		\includegraphics[width=0.999\linewidth]{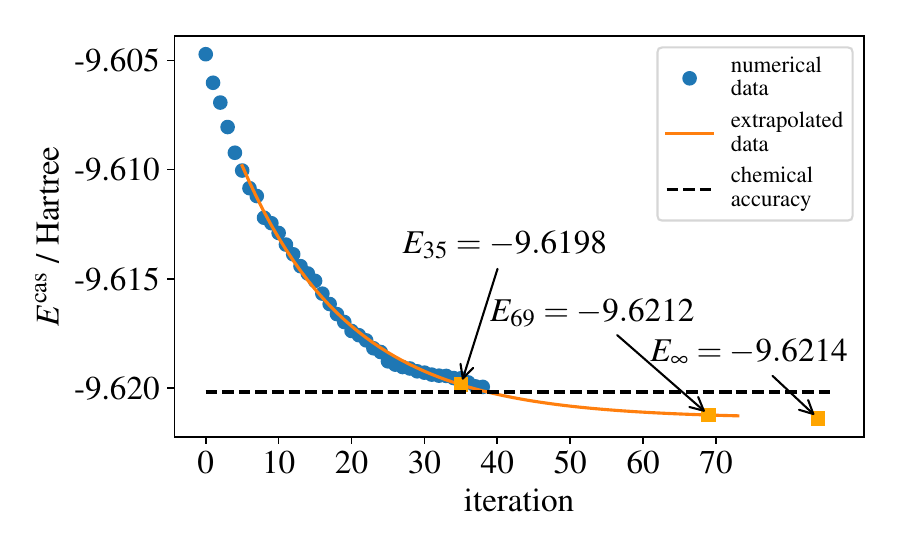}
	\end{subfigure}
	\vspace{-0.1cm}
	\caption{(a): Energy convergence curve of $\mathrm{O_3}$ with active space size (6,6).  (b)-(f) are the extrapolated energy convergence curve at (b): $d=1.43$, (c): $d=1.48$, (d): $d=1.53$, (e): $d=1.58$, and (f) $d=1.63$~\AA. The first highlighted point of each figure corresponds to the iteration and energy by setting the value of $E^{(i)} - E^{(i+1)}$ from~\autoref{eq: energy extrapolation 2} to $1.6 \times 10^{-3}$ Hartree. The second highlighted point gives the iteration number and the expected energy when setting the value of $E^{(i)} - E^{(i+1)} = 1.6 \times 10^{-4}$ Hartree. The last highlighted point represents the expected energy with infinite iterations from the extrapolation. }
	\label{fig: convergence curve and extrapolated energy for o3 with cas (6,6)}
\end{figure*}

\begin{table*}[t]
	\centering
	\caption{\label{tab:table1}Number of optimization parameters}
	\begin{tabularx}{\textwidth} { 
			| >{\centering\arraybackslash}X 
			| >{\centering\arraybackslash}X 
			| >{\centering\arraybackslash}X 
			| >{\centering\arraybackslash}X
			| >{\centering\arraybackslash}X
			| >{\centering\arraybackslash}X
			| >{\centering\arraybackslash}X
			| >{\centering\arraybackslash}X
			| >{\centering\arraybackslash}X
			| >{\centering\arraybackslash}X | }
		\hline \hline
		\multirow{2}{*}{} &
		\multicolumn{3}{c|}{CAS(2,2)}&\multicolumn{3}{c|}{CAS(4,4)}&\multicolumn{3}{c|}{CAS(6,6)}\\  
		\cline{2-10}     &$\mathrm{O_3}$&$\mathrm{Li_4}$&$\mathrm{Cr_2}$&$\mathrm{O_3}$&$\mathrm{Li_4}$&$\mathrm{Cr_2}$&$\mathrm{O_3}$&$\mathrm{Li_4}$&$\mathrm{Cr_2}$\\
		\hline
		QCC & 1 & 1 & 1 & 1--4 & 1--2 & 3 & 24--75 & 3 & 10--12 \\
		UCCSD & 3 & 3 & 3 & 26 & 26 & 26 & 117 & 117 & 117\\
		\hline \hline
	\end{tabularx}
\end{table*}

\begin{table*}
	\centering
	\caption{\label{tab:table2}Hardware parameters}
	\begin{tabularx}{\textwidth}{| >{\centering\arraybackslash}X 
			| >{\centering\arraybackslash}X 
			| >{\centering\arraybackslash}X 
			| >{\centering\arraybackslash}X
			|}
		\hline \hline
		& qubit type  & single-qubit gate infidelity & two-qubit gate infidelity \\ \hline
		IBM Kolkata & superconducting qubits   & 2.199e-4  &7.743e-3 \\
		Quantinuum H1-1 & trapped-ion qubits & 4e-5 & 2e-3 \\
		\hline \hline
	\end{tabularx}
\end{table*}
For CAS(6,6), we refrained from optimizing the UCCSD ansatz due to its high parameter counts~(will be specified in~\autoref{subsec: parameter number count}). QCC, on the other hand, exhibited accurate results under chemical accuracy for $\mathrm{Li_4}$ and $\mathrm{Cr_2}$. However, occasional convergence issues were observed for $\mathrm{O_3}$  within 40 iterations. To address these challenges, we implemented an extrapolation method as suggested in~\cite{ryabinkin_iterative_2020}. The energy difference exhibits an exponential decay, as illustrated in~\autoref{fig: (6,6) o3 convergence all}. Consequently, a logarithmic relationship is assumed:
\begin{subequations}
	\begin{align}
		\log (E^{(i)} - E_{0}) &= ai+b \label{eq: energy extrapolation 1}\\
		\log (E^{(i)} - E^{(i-1)}) &= ai+(b+\log (1-10^{-a})),\label{eq: energy extrapolation 2}
	\end{align}
\end{subequations}
where the parameters $a$ and $b$ are obtained through fitting.
The estimated value $E_{0}^\text{estimated}$ is obtained by solving this fitting problem. The first five iterations were discarded, and the subsequent 35 iterations were employed to address the fitting problem. The results are presented in~\autoref{fig: convergence curve and extrapolated energy for o3 with cas (6,6)}. For $d={1.43}-{1.63}$~\AA, all estimated energies fall within chemical accuracy. By imposing the condition $E^{(i)} - E_{\text{exact}} < 1.6 \times 10^{-4} $ Hartree and identifying the necessary iteration $i$, the energy within chemical accuracy is successfully achieved as well.

These findings support the idea that under the condition that available resources are limited, it may not be necessary to execute the QCC ansatz until the energy difference is smaller than a specific threshold or reaches the maximum iteration limit. Instead, it is already sufficient to run the optimization for a few iterations and subsequently perform extrapolation.

\subsection{Parameter number count}
\label{subsec: parameter number count}

The determination of parameter numbers in the UCCSD ansatz follows a straightforward pattern. For an active space of size (2,2), there are 2 single excitations and 1 double excitation, so 3 parameters are needed. Expanding the active space to (4,4) results in 8 single excitations and 18 double excitations, which means a total of 26 parameters. In the case of an active space with size (6,6), the count increases to 18 types of single excitations and 99 double excitations, leading to 117 parameters. Consequently, as the active space size increases, classical optimization becomes exponentially more challenging.

In contrast, the enhanced QCC ansatz, as illustrated in \autoref{tab:table1}, exhibits notable efficiency. Specifically, for an active space size of (2,2), the energy for $\mathrm{O_3}$, $\mathrm{Li_4}$ and $\mathrm{Cr_2}$ falls below chemical accuracy after just one iteration, equivalent to one Pauli string time evolution gate. With an active size of (4,4), the high dimensionality of the UCCSD ansatz occasionally hampers the optimizer from finding the ground state, as evident in \autoref{fig:pes_total_table}. However, with the enhanced QCC ansatz, chemical accuracy is achieved within a maximum of 4 layers of Pauli string time evolution gates, and for each iteration, only one parameter requires optimization. This stark contrast underscores the efficiency and computational advantages of the enhanced QCC ansatz over the UCCSD ansatz in quantum chemistry calculations.

\section{Experimental results on quantum hardware }
\label{sec: experimental results}

Our experiments were conducted on two distinct quantum hardware platforms: the superconducting-based IBM Kolkata quantum computer~\cite{ibm-kolkata} and the trapped-ion-based Quantinuum H1-1 quantum computer~\cite{quantinuum-h1-1}. The hardware parameters for both devices are outlined in~\autoref{tab:table2}.
Due to constraints such as the number of jobs we can submit and the total runtime available, we chose to run the circuit using classically optimized parameters instead of conducting the entire VQE optimization loop on the quantum computers. The shot number was consistently set to $10^{4}$ for both devices during the experimental runs.
\begin{figure}[ht]
	\centering
	\begin{subfigure}[ht]{0.995\linewidth}
		\caption{}
		\label{fig: circ}
		\includegraphics[width=\linewidth]{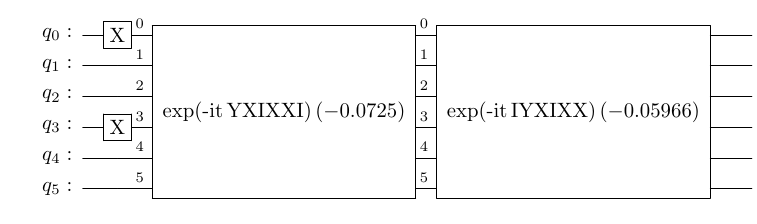}   
	\end{subfigure}
	\begin{subfigure}[ht]{0.995\linewidth}
		\caption{}  
		\label{fig: ibm run energy}  
		\includegraphics[width=\linewidth]{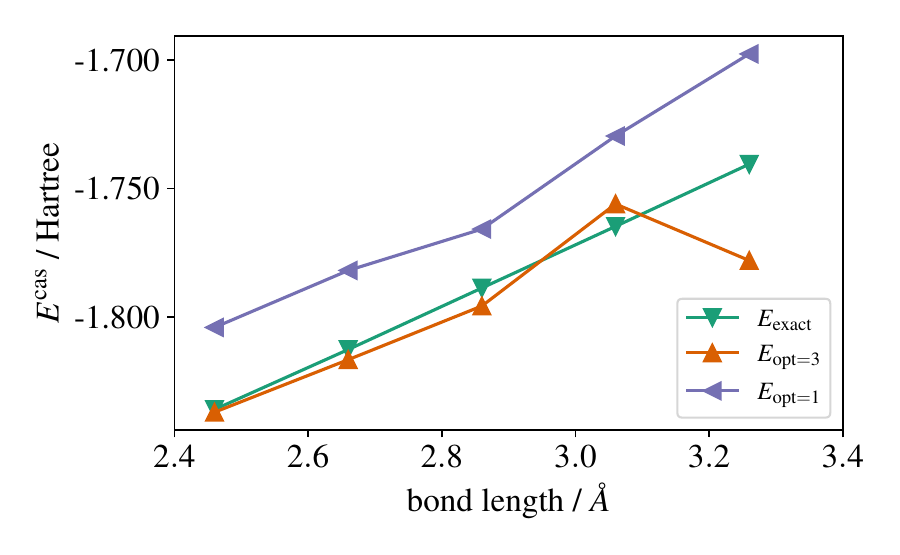}
	\end{subfigure}
	\caption{(a) Circuit with two layers of Pauli time evolution gates for $\mathrm{Li_4}$ with CAS(4,4). (b) The active space energy of $\mathrm{Li}_4$ measured on IBM Kolkata quantum computer with circuit shown in (a).}  
	\label{fig: ibm run with circuit}  
\end{figure}

\begin{figure}[h!]
	\centering
	\includegraphics[width=\linewidth]{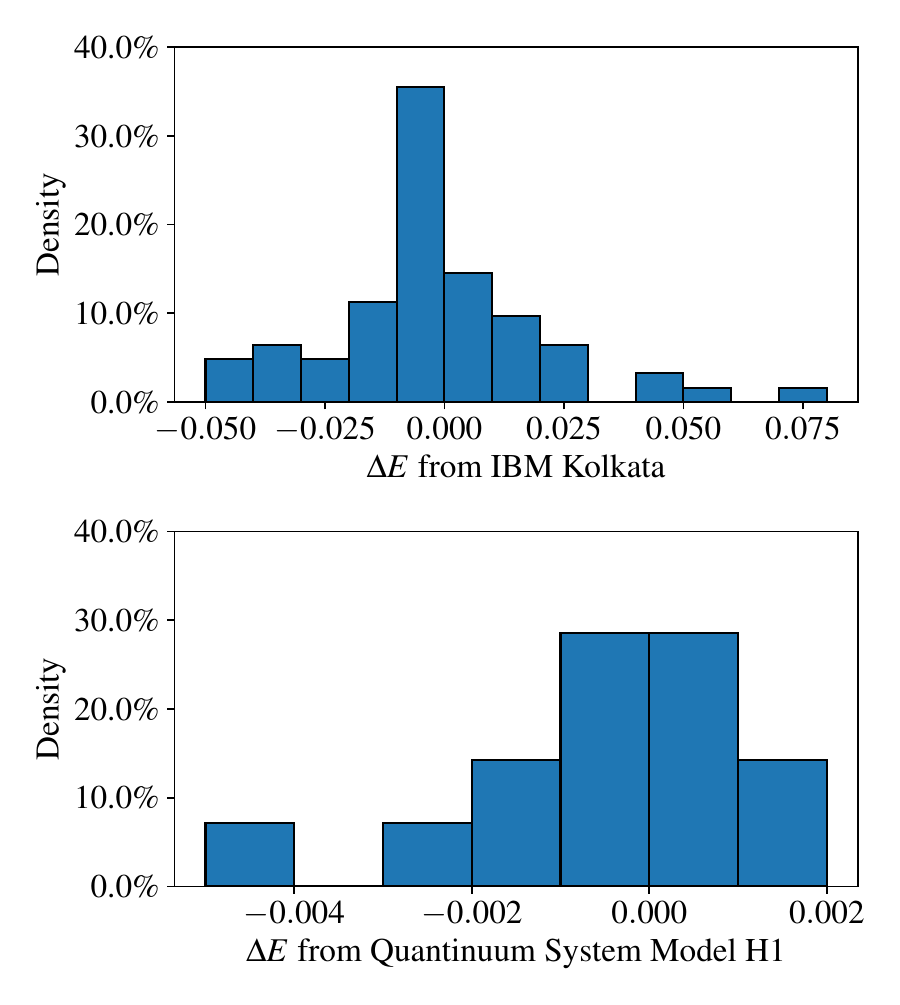}
	\caption{The distribution of the energy difference between the numerical energy $E_\mathrm{num}$ and the measured energy $E_\mathrm{exp}$ for each commuting group of the $\mathrm{Li}_4$ Hamiltonian on IBM Kolkata machine and Quantinuum H1-1 machine.}
	\label{fig:ibm_run_energy_diff}
\end{figure}

We took $\mathrm{Li_4}$ with CAS(4,4) as a representative example. Due to constraints on the coherence time of the qubits, we limited our application to the first two layers of Pauli string time evolution gates, as illustrated in~\autoref{fig: circ}, because the numerical study in last section has shown that two layers of Pauli string time evolution gates are sufficient to lower the energy to chemical accuracy level. To improve the efficiency of the measurement process, we grouped qubit-wise commuting Pauli strings in the Hamiltonian, allowing for simultaneous measurements~\cite{rubin_application_2018, hamamura_efficient_2020}. 

We performed two sets of experiments on the IBM Kolkata platform, one with the optimization level set to 1, and the other set to 3~\cite{qiskit-opt-level}. In each set, we measured the active space energy for bond length ranging from 2.46 to 3.26~\AA. The corresponding energy is plotted in~\autoref{fig: ibm run energy}.
With the optimization level set to 1, the energy trend aligns with the numerically simulated energy, albeit with an offset of approximately $0.4 $ Hartree, which is two magnitudes higher than chemical accuracy. Conversely, with the optimization level set to 3, the discrepancy between the experimental and numerical results diminished across most bond lengths, leaving an offset in the order of $10^{-2}$ Hartree.

On Quantinuum H1-1 quantum computer, we assessed the active space energy of $\mathrm{Li_4}$ at bond length of 2.66~\AA~with optimization level set to 2~\cite{pytket-opt-level, tket-software-overview-paper}. The total energy deviation $E_\mathrm{num} - E_\mathrm{exp} $ is 0.0067 Hartree.

Additionally, the energy difference between numerically simulated and experimentally measured energies for each commuting group under the highest optimization level of both the IBM Kolkata machine and Quantinuum H1-1 machine is displayed in ~\autoref{fig:ibm_run_energy_diff}. The errors within each commuting group fluctuate within the range of $-0.05\textup{--}0.08$ Hartree for IBM Kolkata and $-0.005\textup{--}0.002$ Hartree for Quantinuum H1-1. In many cases, although not universally, these errors tend to cancel each other out. This observation suggests that by measuring the same circuit a sufficiently large number of times on a NISQ device, we can obtain results with a decent level of accuracy.

\section{Conclusion}
\label{sec: conclusion}

In this study, we introduced a Variational Quantum Eigensolver ansatz based on the Qubit Coupled Cluster ansatz. We assessed its performance both through numerical simulations and experimental trials on quantum computers. Numerical simulations showcased the efficiency of our enhanced QCC ansatz, revealing a balance of low circuit depth and parameter counts, and high accuracy in handling both weakly correlated and strongly correlated systems.
Experimental results obtained from quantum hardware experiments demonstrated that, despite inherent noise and other hardware limitations, our proposed ansatz achieved energy measurements at a near-chemical accuracy level. This outcome underscored the robustness and practicality of our approach in real-world quantum computing settings.

In contrast to the original QCC ansatz, which utilized Pauli string time evolution gates that are in general not particle number conserving for energy gradient descent and single-qubit rotation gates for symmetry restoration, our modified approach eliminated the need for single-qubit rotation gates. This was achieved by initiating and maintaining computations within the Hilbert space section with the correct symmetry. 
Consequently, this reduction in gate layers and required parameters not only accelerates the computational process but also raises intriguing questions.
On the one hand, the symmetry-breaking and restoring approach may yield shortcuts, accelerating the search for the ground state and ground state energy. On the other hand, it extends the problem space, introducing the possibility of multiple optimal solution paths for a given problem. This opens topics for further research, especially in the comparative analysis of symmetry-breaking and restoring methods against symmetry-conserving methods. Future investigations in this direction will deepen our understanding and contribute to the refinement of quantum algorithms for chemical systems.

\begin{acknowledgements}
	\label{sec: acknowledgements}
	
	This work was funded by the BMW Group. We thank Edwin Knobbe from BMW Group Battery Comptence Centre for insightful discussions and  continuous support of the project. We acknowledge the use of IBM Quantum services. The views expressed are those of the authors, and do not reflect the official policy or position of IBM or the IBM Quantum team. In addition, we thank Quantinuum Nexus team for providing experiment support and compute access to Quantinuum H1-1.
\end{acknowledgements}
\bibliography{references}

\begin{thebibliography}{45}%
\makeatletter
\providecommand \@ifxundefined [1]{%
 \@ifx{#1\undefined}
}%
\providecommand \@ifnum [1]{%
 \ifnum #1\expandafter \@firstoftwo
 \else \expandafter \@secondoftwo
 \fi
}%
\providecommand \@ifx [1]{%
 \ifx #1\expandafter \@firstoftwo
 \else \expandafter \@secondoftwo
 \fi
}%
\providecommand \natexlab [1]{#1}%
\providecommand \enquote  [1]{``#1''}%
\providecommand \bibnamefont  [1]{#1}%
\providecommand \bibfnamefont [1]{#1}%
\providecommand \citenamefont [1]{#1}%
\providecommand \href@noop [0]{\@secondoftwo}%
\providecommand \href [0]{\begingroup \@sanitize@url \@href}%
\providecommand \@href[1]{\@@startlink{#1}\@@href}%
\providecommand \@@href[1]{\endgroup#1\@@endlink}%
\providecommand \@sanitize@url [0]{\catcode `\\12\catcode `\$12\catcode
  `\&12\catcode `\#12\catcode `\^12\catcode `\_12\catcode `\%12\relax}%
\providecommand \@@startlink[1]{}%
\providecommand \@@endlink[0]{}%
\providecommand \url  [0]{\begingroup\@sanitize@url \@url }%
\providecommand \@url [1]{\endgroup\@href {#1}{\urlprefix }}%
\providecommand \urlprefix  [0]{URL }%
\providecommand \Eprint [0]{\href }%
\providecommand \doibase [0]{https://doi.org/}%
\providecommand \selectlanguage [0]{\@gobble}%
\providecommand \bibinfo  [0]{\@secondoftwo}%
\providecommand \bibfield  [0]{\@secondoftwo}%
\providecommand \translation [1]{[#1]}%
\providecommand \BibitemOpen [0]{}%
\providecommand \bibitemStop [0]{}%
\providecommand \bibitemNoStop [0]{.\EOS\space}%
\providecommand \EOS [0]{\spacefactor3000\relax}%
\providecommand \BibitemShut  [1]{\csname bibitem#1\endcsname}%
\let\auto@bib@innerbib\@empty
\bibitem [{\citenamefont {Peruzzo}\ \emph {et~al.}(2014)\citenamefont
  {Peruzzo}, \citenamefont {McClean}, \citenamefont {Shadbolt}, \citenamefont
  {Yung}, \citenamefont {Zhou}, \citenamefont {Love}, \citenamefont
  {Aspuru-Guzik},\ and\ \citenamefont {O'Brien}}]{peruzzo_variational_2014}%
  \BibitemOpen
  \bibfield  {author} {\bibinfo {author} {\bibfnamefont {A.}~\bibnamefont
  {Peruzzo}}, \bibinfo {author} {\bibfnamefont {J.}~\bibnamefont {McClean}},
  \bibinfo {author} {\bibfnamefont {P.}~\bibnamefont {Shadbolt}}, \bibinfo
  {author} {\bibfnamefont {M.-H.}\ \bibnamefont {Yung}}, \bibinfo {author}
  {\bibfnamefont {X.-Q.}\ \bibnamefont {Zhou}}, \bibinfo {author}
  {\bibfnamefont {P.~J.}\ \bibnamefont {Love}}, \bibinfo {author}
  {\bibfnamefont {A.}~\bibnamefont {Aspuru-Guzik}},\ and\ \bibinfo {author}
  {\bibfnamefont {J.~L.}\ \bibnamefont {O'Brien}},\ }\bibfield  {title}
  {\bibinfo {title} {A variational eigenvalue solver on a photonic quantum
  processor},\ }\href {https://doi.org/10.1038/ncomms5213} {\bibfield
  {journal} {\bibinfo  {journal} {Nat. Commun.}\ }\textbf {\bibinfo {volume}
  {5}},\ \bibinfo {pages} {4213} (\bibinfo {year} {2014})}\BibitemShut
  {NoStop}%
\bibitem [{\citenamefont {Tilly}\ \emph {et~al.}(2022)\citenamefont {Tilly},
  \citenamefont {Chen}, \citenamefont {Cao}, \citenamefont {Picozzi},
  \citenamefont {Setia}, \citenamefont {Li}, \citenamefont {Grant},
  \citenamefont {Wossnig}, \citenamefont {Rungger}, \citenamefont {Booth},\
  and\ \citenamefont {Tennyson}}]{tilly_variational_2022}%
  \BibitemOpen
  \bibfield  {author} {\bibinfo {author} {\bibfnamefont {J.}~\bibnamefont
  {Tilly}}, \bibinfo {author} {\bibfnamefont {H.}~\bibnamefont {Chen}},
  \bibinfo {author} {\bibfnamefont {S.}~\bibnamefont {Cao}}, \bibinfo {author}
  {\bibfnamefont {D.}~\bibnamefont {Picozzi}}, \bibinfo {author} {\bibfnamefont
  {K.}~\bibnamefont {Setia}}, \bibinfo {author} {\bibfnamefont
  {Y.}~\bibnamefont {Li}}, \bibinfo {author} {\bibfnamefont {E.}~\bibnamefont
  {Grant}}, \bibinfo {author} {\bibfnamefont {L.}~\bibnamefont {Wossnig}},
  \bibinfo {author} {\bibfnamefont {I.}~\bibnamefont {Rungger}}, \bibinfo
  {author} {\bibfnamefont {G.~H.}\ \bibnamefont {Booth}},\ and\ \bibinfo
  {author} {\bibfnamefont {J.}~\bibnamefont {Tennyson}},\ }\bibfield  {title}
  {\bibinfo {title} {The {V}ariational {Q}uantum {E}igensolver: {A} review of
  methods and best practices},\ }\href
  {https://doi.org/10.1016/j.physrep.2022.08.003} {\bibfield  {journal}
  {\bibinfo  {journal} {Phys. Rep.}\ }\textbf {\bibinfo {volume} {986}},\
  \bibinfo {pages} {1} (\bibinfo {year} {2022})}\BibitemShut {NoStop}%
\bibitem [{\citenamefont {Sharma}\ \emph {et~al.}(2020)\citenamefont {Sharma},
  \citenamefont {Khatri}, \citenamefont {Cerezo},\ and\ \citenamefont
  {Coles}}]{sharma_noise_2020}%
  \BibitemOpen
  \bibfield  {author} {\bibinfo {author} {\bibfnamefont {K.}~\bibnamefont
  {Sharma}}, \bibinfo {author} {\bibfnamefont {S.}~\bibnamefont {Khatri}},
  \bibinfo {author} {\bibfnamefont {M.}~\bibnamefont {Cerezo}},\ and\ \bibinfo
  {author} {\bibfnamefont {P.~J.}\ \bibnamefont {Coles}},\ }\bibfield  {title}
  {\bibinfo {title} {Noise resilience of variational quantum compiling},\
  }\href {https://doi.org/10.1088/1367-2630/ab784c} {\bibfield  {journal}
  {\bibinfo  {journal} {New J. Phys}\ }\textbf {\bibinfo {volume} {22}},\
  \bibinfo {pages} {043006} (\bibinfo {year} {2020})}\BibitemShut {NoStop}%
\bibitem [{\citenamefont {Fontana}\ \emph {et~al.}(2021)\citenamefont
  {Fontana}, \citenamefont {Fitzpatrick}, \citenamefont {Ramo}, \citenamefont
  {Duncan},\ and\ \citenamefont {Rungger}}]{fontana_evaluating_2021}%
  \BibitemOpen
  \bibfield  {author} {\bibinfo {author} {\bibfnamefont {E.}~\bibnamefont
  {Fontana}}, \bibinfo {author} {\bibfnamefont {N.}~\bibnamefont
  {Fitzpatrick}}, \bibinfo {author} {\bibfnamefont {D.~M.}\ \bibnamefont
  {Ramo}}, \bibinfo {author} {\bibfnamefont {R.}~\bibnamefont {Duncan}},\ and\
  \bibinfo {author} {\bibfnamefont {I.}~\bibnamefont {Rungger}},\ }\bibfield
  {title} {\bibinfo {title} {Evaluating the noise resilience of variational
  quantum algorithms},\ }\href {https://doi.org/10.1103/PhysRevA.104.022403}
  {\bibfield  {journal} {\bibinfo  {journal} {Phys. Rev. A}\ }\textbf {\bibinfo
  {volume} {104}},\ \bibinfo {pages} {022403} (\bibinfo {year}
  {2021})}\BibitemShut {NoStop}%
\bibitem [{\citenamefont {Huggins}\ \emph {et~al.}(2021)\citenamefont
  {Huggins}, \citenamefont {McClean}, \citenamefont {Rubin}, \citenamefont
  {Jiang}, \citenamefont {Wiebe}, \citenamefont {Whaley},\ and\ \citenamefont
  {Babbush}}]{huggins_efficient_2021}%
  \BibitemOpen
  \bibfield  {author} {\bibinfo {author} {\bibfnamefont {W.~J.}\ \bibnamefont
  {Huggins}}, \bibinfo {author} {\bibfnamefont {J.~R.}\ \bibnamefont
  {McClean}}, \bibinfo {author} {\bibfnamefont {N.~C.}\ \bibnamefont {Rubin}},
  \bibinfo {author} {\bibfnamefont {Z.}~\bibnamefont {Jiang}}, \bibinfo
  {author} {\bibfnamefont {N.}~\bibnamefont {Wiebe}}, \bibinfo {author}
  {\bibfnamefont {K.~B.}\ \bibnamefont {Whaley}},\ and\ \bibinfo {author}
  {\bibfnamefont {R.}~\bibnamefont {Babbush}},\ }\bibfield  {title} {\bibinfo
  {title} {Efficient and noise resilient measurements for quantum chemistry on
  near-term quantum computers},\ }\href
  {https://doi.org/10.1038/s41534-020-00341-7} {\bibfield  {journal} {\bibinfo
  {journal} {npj Quantum Information}\ }\textbf {\bibinfo {volume} {7}},\
  \bibinfo {pages} {1} (\bibinfo {year} {2021})}\BibitemShut {NoStop}%
\bibitem [{\citenamefont {Bartlett}\ \emph {et~al.}(1989)\citenamefont
  {Bartlett}, \citenamefont {Kucharski},\ and\ \citenamefont
  {Noga}}]{bartlett_alternative_1989}%
  \BibitemOpen
  \bibfield  {author} {\bibinfo {author} {\bibfnamefont {R.~J.}\ \bibnamefont
  {Bartlett}}, \bibinfo {author} {\bibfnamefont {S.~A.}\ \bibnamefont
  {Kucharski}},\ and\ \bibinfo {author} {\bibfnamefont {J.}~\bibnamefont
  {Noga}},\ }\bibfield  {title} {\bibinfo {title} {Alternative coupled-cluster
  ansätze {II}. {The} unitary coupled-cluster method},\ }\href
  {https://doi.org/10.1016/S0009-2614(89)87372-5} {\bibfield  {journal}
  {\bibinfo  {journal} {Chem. Phys. Lett.}\ }\textbf {\bibinfo {volume}
  {155}},\ \bibinfo {pages} {133} (\bibinfo {year} {1989})}\BibitemShut
  {NoStop}%
\bibitem [{\citenamefont {Romero}\ \emph {et~al.}(2018)\citenamefont {Romero},
  \citenamefont {Babbush}, \citenamefont {McClean}, \citenamefont {Hempel},
  \citenamefont {Love},\ and\ \citenamefont
  {Aspuru-Guzik}}]{romero_strategies_2018}%
  \BibitemOpen
  \bibfield  {author} {\bibinfo {author} {\bibfnamefont {J.}~\bibnamefont
  {Romero}}, \bibinfo {author} {\bibfnamefont {R.}~\bibnamefont {Babbush}},
  \bibinfo {author} {\bibfnamefont {J.~R.}\ \bibnamefont {McClean}}, \bibinfo
  {author} {\bibfnamefont {C.}~\bibnamefont {Hempel}}, \bibinfo {author}
  {\bibfnamefont {P.~J.}\ \bibnamefont {Love}},\ and\ \bibinfo {author}
  {\bibfnamefont {A.}~\bibnamefont {Aspuru-Guzik}},\ }\bibfield  {title}
  {\bibinfo {title} {Strategies for quantum computing molecular energies using
  the unitary coupled cluster ansatz},\ }\href
  {https://doi.org/10.1088/2058-9565/aad3e4} {\bibfield  {journal} {\bibinfo
  {journal} {Quantum Sci. Technol.}\ }\textbf {\bibinfo {volume} {4}},\
  \bibinfo {pages} {014008} (\bibinfo {year} {2018})}\BibitemShut {NoStop}%
\bibitem [{\citenamefont {Filip}\ and\ \citenamefont
  {Thom}(2020)}]{filip_stochastic_2020}%
  \BibitemOpen
  \bibfield  {author} {\bibinfo {author} {\bibfnamefont {M.~A.}\ \bibnamefont
  {Filip}}\ and\ \bibinfo {author} {\bibfnamefont {A.~J.~W.}\ \bibnamefont
  {Thom}},\ }\bibfield  {title} {\bibinfo {title} {A stochastic approach to
  unitary coupled cluster},\ }\href {https://doi.org/10.1063/5.0026141}
  {\bibfield  {journal} {\bibinfo  {journal} {J. Chem. Phys.}\ }\textbf
  {\bibinfo {volume} {153}},\ \bibinfo {pages} {214106} (\bibinfo {year}
  {2020})}\BibitemShut {NoStop}%
\bibitem [{\citenamefont {Metcalf}\ \emph {et~al.}(2020)\citenamefont
  {Metcalf}, \citenamefont {Bauman}, \citenamefont {Kowalski},\ and\
  \citenamefont {de~Jong}}]{metcalf_resource-efficient_2020}%
  \BibitemOpen
  \bibfield  {author} {\bibinfo {author} {\bibfnamefont {M.}~\bibnamefont
  {Metcalf}}, \bibinfo {author} {\bibfnamefont {N.~P.}\ \bibnamefont {Bauman}},
  \bibinfo {author} {\bibfnamefont {K.}~\bibnamefont {Kowalski}},\ and\
  \bibinfo {author} {\bibfnamefont {W.~A.}\ \bibnamefont {de~Jong}},\
  }\bibfield  {title} {\bibinfo {title} {Resource-{Efficient} {Chemistry} on
  {Quantum} {Computers} with the {Variational} {Quantum} {Eigensolver} and the
  {Double} {Unitary} {Coupled}-{Cluster} {Approach}},\ }\href
  {https://doi.org/10.1021/acs.jctc.0c00421} {\bibfield  {journal} {\bibinfo
  {journal} {J. Chem. Theory Comput.}\ }\textbf {\bibinfo {volume} {16}},\
  \bibinfo {pages} {6165} (\bibinfo {year} {2020})}\BibitemShut {NoStop}%
\bibitem [{\citenamefont {Anand}\ \emph {et~al.}(2022)\citenamefont {Anand},
  \citenamefont {Schleich}, \citenamefont {Alperin-Lea}, \citenamefont
  {Jensen}, \citenamefont {Sim}, \citenamefont {Díaz-Tinoco}, \citenamefont
  {Kottmann}, \citenamefont {Degroote}, \citenamefont {Izmaylov},\ and\
  \citenamefont {Aspuru-Guzik}}]{anand_quantum_2022}%
  \BibitemOpen
  \bibfield  {author} {\bibinfo {author} {\bibfnamefont {A.}~\bibnamefont
  {Anand}}, \bibinfo {author} {\bibfnamefont {P.}~\bibnamefont {Schleich}},
  \bibinfo {author} {\bibfnamefont {S.}~\bibnamefont {Alperin-Lea}}, \bibinfo
  {author} {\bibfnamefont {P.~W.~K.}\ \bibnamefont {Jensen}}, \bibinfo {author}
  {\bibfnamefont {S.}~\bibnamefont {Sim}}, \bibinfo {author} {\bibfnamefont
  {M.}~\bibnamefont {Díaz-Tinoco}}, \bibinfo {author} {\bibfnamefont {J.~S.}\
  \bibnamefont {Kottmann}}, \bibinfo {author} {\bibfnamefont {M.}~\bibnamefont
  {Degroote}}, \bibinfo {author} {\bibfnamefont {A.~F.}\ \bibnamefont
  {Izmaylov}},\ and\ \bibinfo {author} {\bibfnamefont {A.}~\bibnamefont
  {Aspuru-Guzik}},\ }\bibfield  {title} {\bibinfo {title} {A quantum computing
  view on unitary coupled cluster theory},\ }\href
  {https://doi.org/10.1039/D1CS00932J} {\bibfield  {journal} {\bibinfo
  {journal} {Chem. Soc. Rev.}\ }\textbf {\bibinfo {volume} {51}},\ \bibinfo
  {pages} {1659} (\bibinfo {year} {2022})}\BibitemShut {NoStop}%
\bibitem [{\citenamefont {Ryabinkin}\ \emph {et~al.}(2018)\citenamefont
  {Ryabinkin}, \citenamefont {Yen}, \citenamefont {Genin},\ and\ \citenamefont
  {Izmaylov}}]{ryabinkin_qubit_2018}%
  \BibitemOpen
  \bibfield  {author} {\bibinfo {author} {\bibfnamefont {I.~G.}\ \bibnamefont
  {Ryabinkin}}, \bibinfo {author} {\bibfnamefont {T.~C.}\ \bibnamefont {Yen}},
  \bibinfo {author} {\bibfnamefont {S.~N.}\ \bibnamefont {Genin}},\ and\
  \bibinfo {author} {\bibfnamefont {A.~F.}\ \bibnamefont {Izmaylov}},\
  }\bibfield  {title} {\bibinfo {title} {Qubit coupled cluster method: {A}
  systematic approach to quantum chemistry on a quantum computer},\ }\href
  {https://doi.org/10.1021/acs.jctc.8b00932} {\bibfield  {journal} {\bibinfo
  {journal} {J. Chem. Theory Comput.}\ }\textbf {\bibinfo {volume} {14}},\
  \bibinfo {pages} {6317} (\bibinfo {year} {2018})}\BibitemShut {NoStop}%
\bibitem [{\citenamefont {Ryabinkin}\ \emph {et~al.}(2020)\citenamefont
  {Ryabinkin}, \citenamefont {Lang}, \citenamefont {Genin},\ and\ \citenamefont
  {Izmaylov}}]{ryabinkin_iterative_2020}%
  \BibitemOpen
  \bibfield  {author} {\bibinfo {author} {\bibfnamefont {I.~G.}\ \bibnamefont
  {Ryabinkin}}, \bibinfo {author} {\bibfnamefont {R.~A.}\ \bibnamefont {Lang}},
  \bibinfo {author} {\bibfnamefont {S.~N.}\ \bibnamefont {Genin}},\ and\
  \bibinfo {author} {\bibfnamefont {A.~F.}\ \bibnamefont {Izmaylov}},\
  }\bibfield  {title} {\bibinfo {title} {Iterative qubit coupled cluster
  approach with efficient screening of generators},\ }\href
  {https://doi.org/10.1021/acs.jctc.9b01084} {\bibfield  {journal} {\bibinfo
  {journal} {J. Chem. Theory Comput.}\ }\textbf {\bibinfo {volume} {16}},\
  \bibinfo {pages} {1055} (\bibinfo {year} {2020})}\BibitemShut {NoStop}%
\bibitem [{\citenamefont {Yordanov}\ \emph {et~al.}(2021)\citenamefont
  {Yordanov}, \citenamefont {Armaos}, \citenamefont {Barnes},\ and\
  \citenamefont {Arvidsson-Shukur}}]{yordanov_qubit-excitation-based_2021}%
  \BibitemOpen
  \bibfield  {author} {\bibinfo {author} {\bibfnamefont {Y.~S.}\ \bibnamefont
  {Yordanov}}, \bibinfo {author} {\bibfnamefont {V.}~\bibnamefont {Armaos}},
  \bibinfo {author} {\bibfnamefont {C.~H.~W.}\ \bibnamefont {Barnes}},\ and\
  \bibinfo {author} {\bibfnamefont {D.~R.~M.}\ \bibnamefont
  {Arvidsson-Shukur}},\ }\bibfield  {title} {\bibinfo {title}
  {Qubit-excitation-based adaptive variational quantum eigensolver},\ }\href
  {https://doi.org/10.1038/s42005-021-00730-0} {\bibfield  {journal} {\bibinfo
  {journal} {Commun. Phys.}\ }\textbf {\bibinfo {volume} {4}},\ \bibinfo
  {pages} {228} (\bibinfo {year} {2021})}\BibitemShut {NoStop}%
\bibitem [{\citenamefont {Kandala}\ \emph {et~al.}(2017)\citenamefont
  {Kandala}, \citenamefont {Mezzacapo}, \citenamefont {Temme}, \citenamefont
  {Takita}, \citenamefont {Brink}, \citenamefont {Chow},\ and\ \citenamefont
  {Gambetta}}]{kandala_hardware-efficient_2017}%
  \BibitemOpen
  \bibfield  {author} {\bibinfo {author} {\bibfnamefont {A.}~\bibnamefont
  {Kandala}}, \bibinfo {author} {\bibfnamefont {A.}~\bibnamefont {Mezzacapo}},
  \bibinfo {author} {\bibfnamefont {K.}~\bibnamefont {Temme}}, \bibinfo
  {author} {\bibfnamefont {M.}~\bibnamefont {Takita}}, \bibinfo {author}
  {\bibfnamefont {M.}~\bibnamefont {Brink}}, \bibinfo {author} {\bibfnamefont
  {J.~M.}\ \bibnamefont {Chow}},\ and\ \bibinfo {author} {\bibfnamefont
  {J.~M.}\ \bibnamefont {Gambetta}},\ }\bibfield  {title} {\bibinfo {title}
  {Hardware-efficient variational quantum eigensolver for small molecules and
  quantum magnets},\ }\href {https://doi.org/10.1038/nature23879} {\bibfield
  {journal} {\bibinfo  {journal} {Nature}\ }\textbf {\bibinfo {volume} {549}},\
  \bibinfo {pages} {242} (\bibinfo {year} {2017})}\BibitemShut {NoStop}%
\bibitem [{\citenamefont {Mitarai}\ \emph {et~al.}(2019)\citenamefont
  {Mitarai}, \citenamefont {Yan},\ and\ \citenamefont
  {Fujii}}]{mitarai_generalization_2019}%
  \BibitemOpen
  \bibfield  {author} {\bibinfo {author} {\bibfnamefont {K.}~\bibnamefont
  {Mitarai}}, \bibinfo {author} {\bibfnamefont {T.}~\bibnamefont {Yan}},\ and\
  \bibinfo {author} {\bibfnamefont {K.}~\bibnamefont {Fujii}},\ }\bibfield
  {title} {\bibinfo {title} {Generalization of the {Output} of a {Variational}
  {Quantum} {Eigensolver} by {Parameter} {Interpolation} with a {Low}-depth
  {Ansatz}},\ }\href {https://doi.org/10.1103/PhysRevApplied.11.044087}
  {\bibfield  {journal} {\bibinfo  {journal} {Phys. Rev. Appl.}\ }\textbf
  {\bibinfo {volume} {11}},\ \bibinfo {pages} {044087} (\bibinfo {year}
  {2019})}\BibitemShut {NoStop}%
\bibitem [{\citenamefont {Grimsley}\ \emph {et~al.}(2019)\citenamefont
  {Grimsley}, \citenamefont {Economou}, \citenamefont {Barnes},\ and\
  \citenamefont {Mayhall}}]{grimsley_adaptive_2019}%
  \BibitemOpen
  \bibfield  {author} {\bibinfo {author} {\bibfnamefont {H.~R.}\ \bibnamefont
  {Grimsley}}, \bibinfo {author} {\bibfnamefont {S.~E.}\ \bibnamefont
  {Economou}}, \bibinfo {author} {\bibfnamefont {E.}~\bibnamefont {Barnes}},\
  and\ \bibinfo {author} {\bibfnamefont {N.~J.}\ \bibnamefont {Mayhall}},\
  }\bibfield  {title} {\bibinfo {title} {An adaptive variational algorithm for
  exact molecular simulations on a quantum computer},\ }\href
  {https://doi.org/10.1038/s41467-019-10988-2} {\bibfield  {journal} {\bibinfo
  {journal} {Nat. Commun.}\ }\textbf {\bibinfo {volume} {10}},\ \bibinfo
  {pages} {3007} (\bibinfo {year} {2019})}\BibitemShut {NoStop}%
\bibitem [{\citenamefont {Claudino}\ \emph {et~al.}(2020)\citenamefont
  {Claudino}, \citenamefont {Wright}, \citenamefont {McCaskey},\ and\
  \citenamefont {Humble}}]{claudino_benchmarking_2020}%
  \BibitemOpen
  \bibfield  {author} {\bibinfo {author} {\bibfnamefont {D.}~\bibnamefont
  {Claudino}}, \bibinfo {author} {\bibfnamefont {J.}~\bibnamefont {Wright}},
  \bibinfo {author} {\bibfnamefont {A.~J.}\ \bibnamefont {McCaskey}},\ and\
  \bibinfo {author} {\bibfnamefont {T.~S.}\ \bibnamefont {Humble}},\ }\bibfield
   {title} {\bibinfo {title} {Benchmarking {Adaptive} {Variational} {Quantum}
  {Eigensolvers}},\ }\href
  {https://www.frontiersin.org/articles/10.3389/fchem.2020.606863} {\bibfield
  {journal} {\bibinfo  {journal} {Front. Chem.}\ }\textbf {\bibinfo {volume}
  {8}} (\bibinfo {year} {2020})}\BibitemShut {NoStop}%
\bibitem [{\citenamefont {Zhang}\ \emph {et~al.}(2021)\citenamefont {Zhang},
  \citenamefont {Kyaw}, \citenamefont {Kottmann}, \citenamefont {Degroote},\
  and\ \citenamefont {Aspuru-Guzik}}]{zhang_mutual_2021}%
  \BibitemOpen
  \bibfield  {author} {\bibinfo {author} {\bibfnamefont {Z.-J.}\ \bibnamefont
  {Zhang}}, \bibinfo {author} {\bibfnamefont {T.~H.}\ \bibnamefont {Kyaw}},
  \bibinfo {author} {\bibfnamefont {J.~S.}\ \bibnamefont {Kottmann}}, \bibinfo
  {author} {\bibfnamefont {M.}~\bibnamefont {Degroote}},\ and\ \bibinfo
  {author} {\bibfnamefont {A.}~\bibnamefont {Aspuru-Guzik}},\ }\bibfield
  {title} {\bibinfo {title} {Mutual information-assisted adaptive variational
  quantum eigensolver},\ }\href {https://doi.org/10.1088/2058-9565/abdca4}
  {\bibfield  {journal} {\bibinfo  {journal} {Quantum Sci. Technol.}\ }\textbf
  {\bibinfo {volume} {6}},\ \bibinfo {pages} {035001} (\bibinfo {year}
  {2021})}\BibitemShut {NoStop}%
\bibitem [{\citenamefont {Tang}\ \emph {et~al.}(2021)\citenamefont {Tang},
  \citenamefont {Shkolnikov}, \citenamefont {Barron}, \citenamefont {Grimsley},
  \citenamefont {Mayhall}, \citenamefont {Barnes},\ and\ \citenamefont
  {Economou}}]{tang_qubit-adapt-vqe_2021}%
  \BibitemOpen
  \bibfield  {author} {\bibinfo {author} {\bibfnamefont {H.~L.}\ \bibnamefont
  {Tang}}, \bibinfo {author} {\bibfnamefont {V.}~\bibnamefont {Shkolnikov}},
  \bibinfo {author} {\bibfnamefont {G.~S.}\ \bibnamefont {Barron}}, \bibinfo
  {author} {\bibfnamefont {H.~R.}\ \bibnamefont {Grimsley}}, \bibinfo {author}
  {\bibfnamefont {N.~J.}\ \bibnamefont {Mayhall}}, \bibinfo {author}
  {\bibfnamefont {E.}~\bibnamefont {Barnes}},\ and\ \bibinfo {author}
  {\bibfnamefont {S.~E.}\ \bibnamefont {Economou}},\ }\bibfield  {title}
  {\bibinfo {title} {Qubit-{ADAPT}-{VQE}: An adaptive algorithm for
  constructing hardware-efficient ans\"atze on a quantum processor},\ }\href
  {https://doi.org/10.1103/PRXQuantum.2.020310} {\bibfield  {journal} {\bibinfo
   {journal} {{PRX} Quantum}\ }\textbf {\bibinfo {volume} {2}},\ \bibinfo
  {pages} {020310} (\bibinfo {year} {2021})}\BibitemShut {NoStop}%
\bibitem [{\citenamefont {Grimsley}\ \emph {et~al.}(2023)\citenamefont
  {Grimsley}, \citenamefont {Barron}, \citenamefont {Barnes}, \citenamefont
  {Economou},\ and\ \citenamefont {Mayhall}}]{grimsley_adapt-vqe_2023}%
  \BibitemOpen
  \bibfield  {author} {\bibinfo {author} {\bibfnamefont {H.~R.}\ \bibnamefont
  {Grimsley}}, \bibinfo {author} {\bibfnamefont {G.~S.}\ \bibnamefont
  {Barron}}, \bibinfo {author} {\bibfnamefont {E.}~\bibnamefont {Barnes}},
  \bibinfo {author} {\bibfnamefont {S.~E.}\ \bibnamefont {Economou}},\ and\
  \bibinfo {author} {\bibfnamefont {N.~J.}\ \bibnamefont {Mayhall}},\
  }\bibfield  {title} {\bibinfo {title} {{ADAPT}-{VQE} is insensitive to rough
  parameter landscapes and barren plateaus},\ }\href
  {https://doi.org/10.1038/s41534-023-00681-0} {\bibfield  {journal} {\bibinfo
  {journal} {npj Quantum Inf}\ }\textbf {\bibinfo {volume} {9}},\ \bibinfo
  {pages} {19} (\bibinfo {year} {2023})}\BibitemShut {NoStop}%
\bibitem [{\citenamefont {Crooks}(2019)}]{crooks_gradients_2019}%
  \BibitemOpen
  \bibfield  {author} {\bibinfo {author} {\bibfnamefont {G.~E.}\ \bibnamefont
  {Crooks}},\ }\bibfield  {title} {\bibinfo {title} {Gradients of parameterized
  quantum gates using the parameter-shift rule and gate decomposition},\
  }\bibfield  {journal} {\bibinfo  {journal} {arXiv.1905.13311}\ }\href
  {https://doi.org/10.48550/arXiv.1905.13311} {10.48550/arXiv.1905.13311}
  (\bibinfo {year} {2019})\BibitemShut {NoStop}%
\bibitem [{\citenamefont {Ostaszewski}\ \emph {et~al.}(2021)\citenamefont
  {Ostaszewski}, \citenamefont {Grant},\ and\ \citenamefont
  {Benedetti}}]{ostaszewski_structure_2021}%
  \BibitemOpen
  \bibfield  {author} {\bibinfo {author} {\bibfnamefont {M.}~\bibnamefont
  {Ostaszewski}}, \bibinfo {author} {\bibfnamefont {E.}~\bibnamefont {Grant}},\
  and\ \bibinfo {author} {\bibfnamefont {M.}~\bibnamefont {Benedetti}},\
  }\bibfield  {title} {\bibinfo {title} {Structure optimization for
  parameterized quantum circuits},\ }\href
  {https://doi.org/10.22331/q-2021-01-28-391} {\bibfield  {journal} {\bibinfo
  {journal} {Quantum}\ }\textbf {\bibinfo {volume} {5}},\ \bibinfo {pages}
  {391} (\bibinfo {year} {2021})}\BibitemShut {NoStop}%
\bibitem [{\citenamefont {Izmaylov}\ \emph {et~al.}(2021)\citenamefont
  {Izmaylov}, \citenamefont {Lang},\ and\ \citenamefont
  {Yen}}]{izmaylov_analytic_2021}%
  \BibitemOpen
  \bibfield  {author} {\bibinfo {author} {\bibfnamefont {A.~F.}\ \bibnamefont
  {Izmaylov}}, \bibinfo {author} {\bibfnamefont {R.~A.}\ \bibnamefont {Lang}},\
  and\ \bibinfo {author} {\bibfnamefont {T.-C.}\ \bibnamefont {Yen}},\
  }\bibfield  {title} {\bibinfo {title} {Analytic gradients in variational
  quantum algorithms: {Algebraic} extensions of the parameter-shift rule to
  general unitary transformations},\ }\href
  {https://doi.org/10.1103/PhysRevA.104.062443} {\bibfield  {journal} {\bibinfo
   {journal} {Phys. Rev. A}\ }\textbf {\bibinfo {volume} {104}},\ \bibinfo
  {pages} {062443} (\bibinfo {year} {2021})}\BibitemShut {NoStop}%
\bibitem [{\citenamefont {Tilly}\ \emph {et~al.}(2021)\citenamefont {Tilly},
  \citenamefont {Sriluckshmy}, \citenamefont {Patel}, \citenamefont {Fontana},
  \citenamefont {Rungger}, \citenamefont {Grant}, \citenamefont {Anderson},
  \citenamefont {Tennyson},\ and\ \citenamefont {Booth}}]{tilly_reduced_2021}%
  \BibitemOpen
  \bibfield  {author} {\bibinfo {author} {\bibfnamefont {J.}~\bibnamefont
  {Tilly}}, \bibinfo {author} {\bibfnamefont {P.~V.}\ \bibnamefont
  {Sriluckshmy}}, \bibinfo {author} {\bibfnamefont {A.}~\bibnamefont {Patel}},
  \bibinfo {author} {\bibfnamefont {E.}~\bibnamefont {Fontana}}, \bibinfo
  {author} {\bibfnamefont {I.}~\bibnamefont {Rungger}}, \bibinfo {author}
  {\bibfnamefont {E.}~\bibnamefont {Grant}}, \bibinfo {author} {\bibfnamefont
  {R.}~\bibnamefont {Anderson}}, \bibinfo {author} {\bibfnamefont
  {J.}~\bibnamefont {Tennyson}},\ and\ \bibinfo {author} {\bibfnamefont
  {G.~H.}\ \bibnamefont {Booth}},\ }\bibfield  {title} {\bibinfo {title}
  {Reduced density matrix sampling: Self-consistent embedding and multiscale
  electronic structure on current generation quantum computers},\ }\href
  {https://doi.org/10.1103/PhysRevResearch.3.033230} {\bibfield  {journal}
  {\bibinfo  {journal} {Phys. Rev. Res.}\ }\textbf {\bibinfo {volume} {3}},\
  \bibinfo {pages} {033230} (\bibinfo {year} {2021})}\BibitemShut {NoStop}%
\bibitem [{\citenamefont {Bentellis}\ \emph {et~al.}(2023)\citenamefont
  {Bentellis}, \citenamefont {Matic-Flierl}, \citenamefont {Mendl},\ and\
  \citenamefont {Lorenz}}]{bentellis_benchmarking_2023}%
  \BibitemOpen
  \bibfield  {author} {\bibinfo {author} {\bibfnamefont {A.}~\bibnamefont
  {Bentellis}}, \bibinfo {author} {\bibfnamefont {A.}~\bibnamefont
  {Matic-Flierl}}, \bibinfo {author} {\bibfnamefont {C.~B.}\ \bibnamefont
  {Mendl}},\ and\ \bibinfo {author} {\bibfnamefont {J.~M.}\ \bibnamefont
  {Lorenz}},\ }\bibfield  {title} {\bibinfo {title} {Benchmarking the
  variational quantum eigensolver using different quantum hardware},\
  }\bibfield  {journal} {\bibinfo  {journal} {arXiv:2305.07092}\ }\href
  {https://doi.org/10.48550/arXiv.2305.07092} {10.48550/arXiv.2305.07092}
  (\bibinfo {year} {2023})\BibitemShut {NoStop}%
\bibitem [{\citenamefont {Helgaker}\ \emph {et~al.}(2014)\citenamefont
  {Helgaker}, \citenamefont {Jorgensen},\ and\ \citenamefont
  {Olsen}}]{helgaker_molecular_2014}%
  \BibitemOpen
  \bibfield  {author} {\bibinfo {author} {\bibfnamefont {T.}~\bibnamefont
  {Helgaker}}, \bibinfo {author} {\bibfnamefont {P.}~\bibnamefont
  {Jorgensen}},\ and\ \bibinfo {author} {\bibfnamefont {J.}~\bibnamefont
  {Olsen}},\ }\href@noop {} {\emph {\bibinfo {title} {{Molecular}
  {Electronic}-{Structure} {Theory}}}}\ (\bibinfo  {publisher} {Wiley},\
  \bibinfo {year} {2014})\BibitemShut {NoStop}%
\bibitem [{\citenamefont {Jordan}\ and\ \citenamefont
  {Wigner}(1928)}]{jordan_uber_1928}%
  \BibitemOpen
  \bibfield  {author} {\bibinfo {author} {\bibfnamefont {P.}~\bibnamefont
  {Jordan}}\ and\ \bibinfo {author} {\bibfnamefont {E.}~\bibnamefont
  {Wigner}},\ }\bibfield  {title} {\bibinfo {title} {\"uber das paulische
  \"aquivalenzverbot},\ }\href {https://doi.org/10.1007/BF01331938} {\bibfield
  {journal} {\bibinfo  {journal} {Zeitschrift für Physik}\ }\textbf {\bibinfo
  {volume} {47}},\ \bibinfo {pages} {631} (\bibinfo {year} {1928})}\BibitemShut
  {NoStop}%
\bibitem [{\citenamefont {Bravyi}\ and\ \citenamefont
  {Kitaev}(2002)}]{bravyi_fermionic_2002}%
  \BibitemOpen
  \bibfield  {author} {\bibinfo {author} {\bibfnamefont {S.~B.}\ \bibnamefont
  {Bravyi}}\ and\ \bibinfo {author} {\bibfnamefont {A.~Y.}\ \bibnamefont
  {Kitaev}},\ }\bibfield  {title} {\bibinfo {title} {Fermionic quantum
  computation},\ }\href {https://doi.org/10.1006/aphy.2002.6254} {\bibfield
  {journal} {\bibinfo  {journal} {Annals of Physics}\ }\textbf {\bibinfo
  {volume} {298}},\ \bibinfo {pages} {210} (\bibinfo {year}
  {2002})}\BibitemShut {NoStop}%
\bibitem [{\citenamefont {Seeley}\ \emph {et~al.}(2012)\citenamefont {Seeley},
  \citenamefont {Richard},\ and\ \citenamefont
  {Love}}]{seeley_bravyi-kitaev_2012}%
  \BibitemOpen
  \bibfield  {author} {\bibinfo {author} {\bibfnamefont {J.~T.}\ \bibnamefont
  {Seeley}}, \bibinfo {author} {\bibfnamefont {M.~J.}\ \bibnamefont
  {Richard}},\ and\ \bibinfo {author} {\bibfnamefont {P.~J.}\ \bibnamefont
  {Love}},\ }\bibfield  {title} {\bibinfo {title} {The {B}ravyi-{K}itaev
  transformation for quantum computation of electronic structure},\ }\href
  {https://doi.org/10.1063/1.4768229} {\bibfield  {journal} {\bibinfo
  {journal} {J. Chem. Phys.}\ }\textbf {\bibinfo {volume} {137}},\ \bibinfo
  {pages} {224109} (\bibinfo {year} {2012})}\BibitemShut {NoStop}%
\bibitem [{\citenamefont {Grimsley}\ \emph {et~al.}(2020)\citenamefont
  {Grimsley}, \citenamefont {Claudino}, \citenamefont {Economou}, \citenamefont
  {Barnes},\ and\ \citenamefont {Mayhall}}]{grimsley_is_2020}%
  \BibitemOpen
  \bibfield  {author} {\bibinfo {author} {\bibfnamefont {H.~R.}\ \bibnamefont
  {Grimsley}}, \bibinfo {author} {\bibfnamefont {D.}~\bibnamefont {Claudino}},
  \bibinfo {author} {\bibfnamefont {S.~E.}\ \bibnamefont {Economou}}, \bibinfo
  {author} {\bibfnamefont {E.}~\bibnamefont {Barnes}},\ and\ \bibinfo {author}
  {\bibfnamefont {N.~J.}\ \bibnamefont {Mayhall}},\ }\bibfield  {title}
  {\bibinfo {title} {Is the {Trotterized} {UCCSD} {Ansatz} chemically
  well-defined?},\ }\href {https://doi.org/10.1021/acs.jctc.9b01083} {\bibfield
   {journal} {\bibinfo  {journal} {J. Chem. Theory Comput.}\ }\textbf {\bibinfo
  {volume} {16}},\ \bibinfo {pages} {1} (\bibinfo {year} {2020})}\BibitemShut
  {NoStop}%
\bibitem [{\citenamefont {Szabo}\ and\ \citenamefont
  {Ostlund}(1996)}]{szabo1996modern}%
  \BibitemOpen
  \bibfield  {author} {\bibinfo {author} {\bibfnamefont {A.}~\bibnamefont
  {Szabo}}\ and\ \bibinfo {author} {\bibfnamefont {N.}~\bibnamefont
  {Ostlund}},\ }\href@noop {} {\emph {\bibinfo {title} {Modern Quantum
  Chemistry: Introduction to Advanced Electronic Structure Theory}}}\ (\bibinfo
   {publisher} {Dover Publications},\ \bibinfo {year} {1996})\ \bibinfo {note}
  {page 40 - 45}\BibitemShut {NoStop}%
\bibitem [{\citenamefont {{McClean}}\ \emph {et~al.}(2018)\citenamefont
  {{McClean}}, \citenamefont {Boixo}, \citenamefont {Smelyanskiy},
  \citenamefont {Babbush},\ and\ \citenamefont {Neven}}]{mcclean_barren_2018}%
  \BibitemOpen
  \bibfield  {author} {\bibinfo {author} {\bibfnamefont {J.~R.}\ \bibnamefont
  {{McClean}}}, \bibinfo {author} {\bibfnamefont {S.}~\bibnamefont {Boixo}},
  \bibinfo {author} {\bibfnamefont {V.~N.}\ \bibnamefont {Smelyanskiy}},
  \bibinfo {author} {\bibfnamefont {R.}~\bibnamefont {Babbush}},\ and\ \bibinfo
  {author} {\bibfnamefont {H.}~\bibnamefont {Neven}},\ }\bibfield  {title}
  {\bibinfo {title} {Barren plateaus in quantum neural network training
  landscapes},\ }\href {https://doi.org/10.1038/s41467-018-07090-4} {\bibfield
  {journal} {\bibinfo  {journal} {Nat. Commun.}\ }\textbf {\bibinfo {volume}
  {9}},\ \bibinfo {pages} {4812} (\bibinfo {year} {2018})}\BibitemShut
  {NoStop}%
\bibitem [{\citenamefont {Wang}\ \emph {et~al.}(2021)\citenamefont {Wang},
  \citenamefont {Fontana}, \citenamefont {Cerezo}, \citenamefont {Sharma},
  \citenamefont {Sone}, \citenamefont {Cincio},\ and\ \citenamefont
  {Coles}}]{wang_noise-induced_2021}%
  \BibitemOpen
  \bibfield  {author} {\bibinfo {author} {\bibfnamefont {S.}~\bibnamefont
  {Wang}}, \bibinfo {author} {\bibfnamefont {E.}~\bibnamefont {Fontana}},
  \bibinfo {author} {\bibfnamefont {M.}~\bibnamefont {Cerezo}}, \bibinfo
  {author} {\bibfnamefont {K.}~\bibnamefont {Sharma}}, \bibinfo {author}
  {\bibfnamefont {A.}~\bibnamefont {Sone}}, \bibinfo {author} {\bibfnamefont
  {L.}~\bibnamefont {Cincio}},\ and\ \bibinfo {author} {\bibfnamefont {P.~J.}\
  \bibnamefont {Coles}},\ }\bibfield  {title} {\bibinfo {title} {Noise-induced
  barren plateaus in variational quantum algorithms},\ }\href
  {https://doi.org/10.1038/s41467-021-27045-6} {\bibfield  {journal} {\bibinfo
  {journal} {Nat. Commun.}\ }\textbf {\bibinfo {volume} {12}},\ \bibinfo
  {pages} {6961} (\bibinfo {year} {2021})}\BibitemShut {NoStop}%
\bibitem [{par()}]{particle-number}%
  \BibitemOpen
  \href@noop {} {}\bibinfo {note} {We use a generalized definition of particle
  number here: it is the expectation value of the number operator $\langle N
  \rangle$, instead of the eigenvalue of the particle number operator $N
  \ket{\psi} = N_e \ket{\psi}$. The generalized definition comes from using the
  trace of one-electron reduced density matrix (1eRDM) as the number of
  electrons, i.e. $\mathrm{Tr}(\mathrm{1eRDM}) = \sum_{i} \bra{\psi}
  a_i^\dagger a_i \ket{\psi} = \bra{\psi} N \ket{\psi}$.}\BibitemShut {Stop}%
\bibitem [{\citenamefont {Rossmannek}\ \emph {et~al.}(2021)\citenamefont
  {Rossmannek}, \citenamefont {Barkoutsos}, \citenamefont {Ollitrault},\ and\
  \citenamefont {Tavernelli}}]{rossmannek_quantum_2021}%
  \BibitemOpen
  \bibfield  {author} {\bibinfo {author} {\bibfnamefont {M.}~\bibnamefont
  {Rossmannek}}, \bibinfo {author} {\bibfnamefont {P.~K.}\ \bibnamefont
  {Barkoutsos}}, \bibinfo {author} {\bibfnamefont {P.~J.}\ \bibnamefont
  {Ollitrault}},\ and\ \bibinfo {author} {\bibfnamefont {I.}~\bibnamefont
  {Tavernelli}},\ }\bibfield  {title} {\bibinfo {title} {Quantum
  {HF}/{DFT}-embedding algorithms for electronic structure calculations:
  {S}caling up to complex molecular systems},\ }\href
  {https://doi.org/10.1063/5.0029536} {\bibfield  {journal} {\bibinfo
  {journal} {J. Chem. Phys.}\ }\textbf {\bibinfo {volume} {154}},\ \bibinfo
  {pages} {114105} (\bibinfo {year} {2021})}\BibitemShut {NoStop}%
\bibitem [{\citenamefont {Dunning}(1989)}]{dunning_gaussian_1989}%
  \BibitemOpen
  \bibfield  {author} {\bibinfo {author} {\bibfnamefont {T.~H.}\ \bibnamefont
  {Dunning}},\ }\bibfield  {title} {\bibinfo {title} {Gaussian basis sets for
  use in correlated molecular calculations. {I}. {The} atoms boron through neon
  and hydrogen},\ }\href {https://doi.org/10.1063/1.456153} {\bibfield
  {journal} {\bibinfo  {journal} {J. Chem. Phys.}\ }\textbf {\bibinfo {volume}
  {90}},\ \bibinfo {pages} {1007} (\bibinfo {year} {1989})}\BibitemShut
  {NoStop}%
\bibitem [{\citenamefont {Sun}\ \emph {et~al.}(2018)\citenamefont {Sun},
  \citenamefont {Berkelbach}, \citenamefont {Blunt}, \citenamefont {Booth},
  \citenamefont {Guo}, \citenamefont {Li}, \citenamefont {Liu}, \citenamefont
  {{McClain}}, \citenamefont {Sayfutyarova}, \citenamefont {Sharma},
  \citenamefont {Wouters},\ and\ \citenamefont {Chan}}]{sun_pyscf_2018}%
  \BibitemOpen
  \bibfield  {author} {\bibinfo {author} {\bibfnamefont {Q.}~\bibnamefont
  {Sun}}, \bibinfo {author} {\bibfnamefont {T.~C.}\ \bibnamefont {Berkelbach}},
  \bibinfo {author} {\bibfnamefont {N.~S.}\ \bibnamefont {Blunt}}, \bibinfo
  {author} {\bibfnamefont {G.~H.}\ \bibnamefont {Booth}}, \bibinfo {author}
  {\bibfnamefont {S.}~\bibnamefont {Guo}}, \bibinfo {author} {\bibfnamefont
  {Z.}~\bibnamefont {Li}}, \bibinfo {author} {\bibfnamefont {J.}~\bibnamefont
  {Liu}}, \bibinfo {author} {\bibfnamefont {J.~D.}\ \bibnamefont {{McClain}}},
  \bibinfo {author} {\bibfnamefont {E.~R.}\ \bibnamefont {Sayfutyarova}},
  \bibinfo {author} {\bibfnamefont {S.}~\bibnamefont {Sharma}}, \bibinfo
  {author} {\bibfnamefont {S.}~\bibnamefont {Wouters}},\ and\ \bibinfo {author}
  {\bibfnamefont {G.~K.-L.}\ \bibnamefont {Chan}},\ }\bibfield  {title}
  {\bibinfo {title} {{PySCF}: the python-based simulations of chemistry
  framework},\ }\href {https://doi.org/10.1002/wcms.1340} {\bibfield  {journal}
  {\bibinfo  {journal} {{WIREs} Comput Mol Sci}\ }\textbf {\bibinfo {volume}
  {8}},\ \bibinfo {pages} {e1340} (\bibinfo {year} {2018})}\BibitemShut
  {NoStop}%
\bibitem [{\citenamefont {{Qiskit contributors}}(2023)}]{Qiskit}%
  \BibitemOpen
  \bibfield  {author} {\bibinfo {author} {\bibnamefont {{Qiskit
  contributors}}},\ }\href {https://doi.org/10.5281/zenodo.2573505} {\bibinfo
  {title} {Qiskit: An open-source framework for quantum computing}} (\bibinfo
  {year} {2023})\BibitemShut {NoStop}%
\bibitem [{ibm()}]{ibm-kolkata}%
  \BibitemOpen
  \href@noop {} {}\bibinfo {note} {IBM Quantum. https://quantum.ibm.com/,
  experiments were conducted on 19th-20th December, 2023.}\BibitemShut {Stop}%
\bibitem [{qua()}]{quantinuum-h1-1}%
  \BibitemOpen
  \href@noop {} {}\bibinfo {note} {Quantinuum H1-1.
  https://www.quantinuum.com/, experiments were conducted on 8th January,
  2024.}\BibitemShut {Stop}%
\bibitem [{\citenamefont {Rubin}\ \emph {et~al.}(2018)\citenamefont {Rubin},
  \citenamefont {Babbush},\ and\ \citenamefont
  {McClean}}]{rubin_application_2018}%
  \BibitemOpen
  \bibfield  {author} {\bibinfo {author} {\bibfnamefont {N.~C.}\ \bibnamefont
  {Rubin}}, \bibinfo {author} {\bibfnamefont {R.}~\bibnamefont {Babbush}},\
  and\ \bibinfo {author} {\bibfnamefont {J.}~\bibnamefont {McClean}},\
  }\bibfield  {title} {\bibinfo {title} {Application of fermionic marginal
  constraints to hybrid quantum algorithms},\ }\href
  {https://doi.org/10.1088/1367-2630/aab919} {\bibfield  {journal} {\bibinfo
  {journal} {New J. Phys.}\ }\textbf {\bibinfo {volume} {20}},\ \bibinfo
  {pages} {053020} (\bibinfo {year} {2018})}\BibitemShut {NoStop}%
\bibitem [{\citenamefont {Hamamura}\ and\ \citenamefont
  {Imamichi}(2020)}]{hamamura_efficient_2020}%
  \BibitemOpen
  \bibfield  {author} {\bibinfo {author} {\bibfnamefont {I.}~\bibnamefont
  {Hamamura}}\ and\ \bibinfo {author} {\bibfnamefont {T.}~\bibnamefont
  {Imamichi}},\ }\bibfield  {title} {\bibinfo {title} {Efficient evaluation of
  quantum observables using entangled measurements},\ }\href
  {https://doi.org/10.1038/s41534-020-0284-2} {\bibfield  {journal} {\bibinfo
  {journal} {npj Quantum Inf}\ }\textbf {\bibinfo {volume} {6}},\ \bibinfo
  {pages} {1} (\bibinfo {year} {2020})}\BibitemShut {NoStop}%
\bibitem [{qis()}]{qiskit-opt-level}%
  \BibitemOpen
  \href@noop {} {}\bibinfo {note} {Qiskit compiler offers four optimization
  levels (0-4) for circuit optimization, ranging from 0 indicating no
  optimization to 3 representing heavy optimization. Similarly, it provides
  four resilience levels (0-4) for error mitigation, with 0 denoting no
  mitigation and 3 indicating heavy mitigation. Throughout the paper, we
  collectively refer to both settings as the optimization level.}\BibitemShut
  {Stop}%
\bibitem [{pyt()}]{pytket-opt-level}%
  \BibitemOpen
  \href@noop {} {}\bibinfo {note} {The circuit to run on the Quantinuum H1
  machine was compiled using the pytket compiler, with optimization levels
  ranging from 0 to 2. Here, 0 signifies low optimization, while 2 corresponds
  to high optimization.}\BibitemShut {Stop}%
\bibitem [{\citenamefont {Sivarajah}\ \emph {et~al.}(2020)\citenamefont
  {Sivarajah}, \citenamefont {Dilkes}, \citenamefont {Cowtan}, \citenamefont
  {Simmons}, \citenamefont {Edgington},\ and\ \citenamefont
  {Duncan}}]{tket-software-overview-paper}%
  \BibitemOpen
  \bibfield  {author} {\bibinfo {author} {\bibfnamefont {S.}~\bibnamefont
  {Sivarajah}}, \bibinfo {author} {\bibfnamefont {S.}~\bibnamefont {Dilkes}},
  \bibinfo {author} {\bibfnamefont {A.}~\bibnamefont {Cowtan}}, \bibinfo
  {author} {\bibfnamefont {W.}~\bibnamefont {Simmons}}, \bibinfo {author}
  {\bibfnamefont {A.}~\bibnamefont {Edgington}},\ and\ \bibinfo {author}
  {\bibfnamefont {R.}~\bibnamefont {Duncan}},\ }\bibfield  {title} {\bibinfo
  {title} {t$\vert$ket⟩: a retargetable compiler for {NISQ} devices},\ }\href
  {https://doi.org/10.1088/2058-9565/ab8e92} {\bibfield  {journal} {\bibinfo
  {journal} {Quantum Sci. Technol.}\ }\textbf {\bibinfo {volume} {6}},\
  \bibinfo {pages} {014003} (\bibinfo {year} {2020})}\BibitemShut {NoStop}%
\end{thebibliography}%
\end{document}